\documentclass[pra,10pt,twocolumn,reprint,superscriptaddress,floatfix,showpacs]{revtex4-2}

\usepackage[utf8]{inputenc}
\usepackage[T1]{fontenc}     
\usepackage[british]{babel}  
\usepackage[sc,osf]{mathpazo}
\usepackage[scaled=0.86]{berasans}  
\usepackage[colorlinks = true,
linkcolor = blue,
urlcolor  = blue,
citecolor = blue,
anchorcolor = blue]{hyperref}  
\usepackage{graphicx} 
\usepackage{epstopdf}
\usepackage{subfig}
\usepackage[babel]{microtype}  
\usepackage{amsmath,amssymb,amsthm,bm,amsfonts,mathrsfs,bbm} 

\usepackage{xspace}  
\usepackage{pgfplots}
\usepackage{xcolor,colortbl}
\def\ba{\begin{equation}}
\def\ea{\end{equation}}
\def\bea{\begin{eqnarray}}
\def\eea{\end{eqnarray}}
\def\ben{\begin{equation*}}
\def\een{\end{equation*}}
\def\bean{\begin{eqnarray*}}
\def\eean{\end{eqnarray*}}
\def\bma{\begin{mathletters}}
\def\ema{\end{mathletters}}
\def\bi{\begin{itemize}}
\def\ei{\end{itemize}}

\newcommand{\be}{\begin{equation}}
\newcommand{\ee}{\end{equation}}

\newcommand{\kommentar}[1]{}

\newcommand{\forget}[1]{}

\begin{document}

\title{Hidden Steering Nonlocality in Quantum Networks}
\author{Kaushiki Mukherjee*}
\email{kaushiki.wbes@gmail.com}
\affiliation{Department of Mathematics, Government Girls' General Degree College, Ekbalpore, Kolkata-700023, India.}
\author{Biswajit Paul}
\email{biswajitpaul4@gmail.com}
\affiliation{Department of Mathematics, Balagarh Bijoykrishna Mahavidyalaya, Balagarh, Dist. - Hooghly-712501, India}
\author{Soma Mandal}
\email{soma2778@gmail.com}
\affiliation{Department of Physics, Government Girls' General Degree College, Ekbalpore, Kolkata-700023, India.}

\begin{abstract}

By combining two objects with no quantum effect one can get an object with quantum effect. Such a phenomenon, often referred to as \textit{activation} has been analyzed for the notion of steering nonlocality. Activation of steering nonlocality is observed for different classes of mixed entangled states in linear network scenarios. Characterization of arbitrary two qubit states, in ambit of steering activation in network scenarios has been provided in this context. Using the notion of reduced steering, instances of steerability activation are also observed in nonlinear network. Present analysis involves three measurement settings scenario(for both trusted and untrusted parties) where steering nonlocality is distinguishable from Bell nonlocality.
\end{abstract}

\maketitle

	
\section{I. Introduction}
Quantum nonlocality is  is an inherent feature of quantum theory\cite{Ein,bel}. It forms the basis
of various information theoretic tasks\cite{cle,may,bru,bar,aci,pir,col,ban}. Presence of entanglement is
a necessary condition for generation of nonlocal correlations, though it is not sufficient due to
existence of local models of some mixed entangled states\cite{wer,wer1,wer2}. Such type of entangled
states are often referred to as \textit{local entangled states}\cite{wer3}.
Procedures involving exploitation of nonlocal correlations from local entangled states are often referred to as \textit{activation scenarios.}\cite{wer4}. Till date, such activation scenarios are classified into three categories: \textit{activation via local filtering}\cite{pop,gis,pal}, \textit{activation by tensoring}\cite{tensor1,ten1,ten2,ten3,ten4} and \textit{activation in quantum networks.}. Any possible combination of mechanisms involved in these three types is also considered as a valid activation procedure.\\

In activation by quantum network scenarios, nonlocality is activated by suitable arrangement of
states(different or identical copies) in a quantum network\cite{sen,cav,caa,woj,klo}.
Speaking of the role of quantum networks in activation, entanglement swapping networks have
emerged as an useful tool for activating nonlocality of states in standard
Bell scenario. In present discussion, utility of these networks will be explored in ambit of activating
nonlocality beyond Bell scenario. \\
In an entanglement swapping network, entanglement is created between two distant parties sharing no direct
common past\cite{zko,zuk,pan}. Apart from its fundamental importance, it is applicable in various quantum applications.
This procedure is also a specific example of quantum teleportation\cite{ben}. \\
The key point of quantum nonlocality activation(Bell-CHSH sense) in entanglement swapping scenario is
that starting from entangled
states(shared between interacting parties) satisfying Bell-CHSH inequality, a Bell-nonlocal state is generated between non interacting parties at the end of the protocol. In \cite{sen,woj,klo} swapping procedure has been framed as a novel example of nonlocality activation in quantum mechanics. Existing research works
have exploited bipartite\cite{sen,woj,klo} and tripartite hidden nonlocality\cite{bis1} in standard Bell scenario using swapping network. Present work will be exploring the utility(if any) of entanglement
swapping network for activation of quantum steering nonlocality. Owing to involvement of
sequential measurements in the network scenario, we will refer activation of steering nonlocality
as \textit{revealing hidden steering nonlocality} in spirit of Popescu\cite{pop}.  \\
Motivated by famous EPR argument\cite{Ein} claiming incompleteness of quantum theory, Schrodinger
first gave the concept of \textit{steering}\cite{str1,str2}. A complete mathematical formalism of such a manifestation of steering was provided
in \cite{jon} where they characterized \textit{steering correlations}. Several criteria
have emerged for detecting steerability of correlations generated from a
given quantum state\cite{red1,red2,red3,red4,zow,red5,san,ver,jev,sjj,costa}. The correlation based
criterion given in \cite{red3}, often referred to as CJWR inequality, is used here for analyzing activation
of steerability. Up to two measurement settings scenario, notions of Bell-CHSH nonlocality and any steering
nonlocality are indistinguishable. So, here we consider CJWR inequality for three measurement settings.
Violation of this symmetric inequality guarantees steerability of the bipartite correlations generated
in the corresponding measurement scenario. Such form of steerability is often referred to as
\textit{$F_3$ steerability}. Using such a symmetric inequality as a detection criterion allows
interchange of the roles of the trusted and the untrusted parties in the operational interpretation
of steering.  \\
Now consider a scenario involving two entangled states($\rho_{AB},\rho_{BC}$,say) such that none of
them violates CJWR inequality for three settings\cite{red3}. Let $\rho_{AB}$ and $\rho_{BC}$ be shared
between three distant parties Alice, Bob and Charlie(say) where Alice and Charlie share no direct common
past. Let $\rho_{AB}$ be shared between Alice and Bob(say) whereas $\rho_{BC}$ be shared between Bob and
Charlie. Let classical communication be allowed between two parties sharing a state. Hence, Alice and
Charlie do not interact. In such a scenario, when the parties perform local operations, will it be
possible to generate a steerable state between the two non interacting parties$?\,$Affirmative result is
obtained when one considers an entanglement swapping network. To be precise, for some outputs of Bob, conditional state shared between the two non interacting parties(Alice and Charlie) turn out to be $F_3$ steerable.\\
After observing hidden steerability for some families of two qubit states in a standard entanglement swapping
network(Fig.\ref{fig1}), a characterization of arbitrary two qubits states is given in this context.
As already mentioned before, CJWR inequality (for three settings) given in \cite{red3} is used as a
detection criterion.
Instance of genuine activation of steering is also observed in the sense that steerable state is obtained
while using unsteerable states in the swapping protocol. Arbitrary two qubit states have also been
characterized in perspective of genuine activation. At this junction it should be pointed out that
the steerable conditional states resulting at the end of the protocol are Bell-local in corresponding
measurement scenario\cite{i3322}. \\
Exploring hidden steerability in three party entanglement swapping scheme, number of parties is then increased. Results of activation are observed in a star network configuration of
entanglement swapping involving non-linear arrangement of four parties under some suitable measurement contexts. \\
Rest of our work is organized as follows. In Sec.\ref{mot1}, we provide the motivation
underlying present discussion. In Sec.\ref{pre1}, we provide with some mathematical preliminaries. Activation of steerability in three party network scenario is
analyzed in Sec.\ref{hid}. In next section, revelation of hidden steerability is then
discussed when number of parties is increased in a non linear
fashion(in Sec.\ref{lin2}). Phenomenon of genuine activation of steering nonlocality is discussed in
Sec.\ref{pur} followed by concluding remarks in Sec.\ref{conc}.
\section{Motivation}\label{mot1}
Steerable correlations are used in various quantum information processing tasks such as
cryptography\cite{revw1,revw2,revw3,revw4,revw5,revw6}, randomness
certification\cite{revw7,revw8,revw9,revw10,revw11}$,\,$channel discrimination\cite{revw12,revw13} and many
others. So any steerable quantum state is considered an useful resource. Though pure entangled states
are best candidate in this context, but these are hardly available. Consequently, mixed entangled states
are used in practical situations all of which are not steerable. From practical perspectives, exploiting steerability from
unsteerable entangled states thus warrants attention. In this context revelation
of hidden steerability from unsteerable quantum states basically motivates present discussion. Choosing network scenario based on entanglement swapping for the activation purpose is further motivated by the fact that steerable correlations can be generated between two non interacting parties once the states involved are subjected to suitable LOCC\cite{hor}. Such nonclassical correlations in turn may be used as a resource in network based quantum information and communication protocols\cite{rest1,rest2,rest3}.
\section{Preliminaries}\label{pre1}
\subsection{Bloch Vector Representation}
Let $\varrho$ denote a two qubit state shared between two parties.
\begin{equation}\label{st4}
\small{\varrho}=\small{\frac{1}{4}(\mathbb{I}_{2\times2}+\vec{\mathfrak{u}}.\vec{\sigma}\otimes \mathbb{I}_2+\mathbb{I}_2\otimes \vec{\mathfrak{v}}.\vec{\sigma}+\sum_{j_1,j_2=1}^{3}w_{j_1j_2}\sigma_{j_1}\otimes\sigma_{j_2})},
\end{equation}
with $\vec{\sigma}$$=$$(\sigma_1,\sigma_2,\sigma_3), $ $\sigma_{j_k}$ denoting Pauli operators along three mutually perpendicular directions($j_k$$=$$1,2,3$). $\vec{\mathfrak{u}}$$=$$(l_1,l_2,l_3)$ and $\vec{\mathfrak{v}}$$=$$(r_1,r_2,r_3)$ stand for the local bloch vectors($\vec{\mathfrak{u}},\vec{\mathfrak{v}}$$\in$$\mathbb{R}^3$) of party $\mathcal{A}$ and $\mathcal{B}$ respectively with $|\vec{\mathfrak{u}}|,|\vec{\mathfrak{v}}|$$\leq$$1$ and $(w_{i,j})_{3\times3}$ denotes the correlation tensor $\mathcal{W}$(a real matrix).
The components $w_{j_1j_2}$ are given by $w_{j_1j_2}$$=$$\textmd{Tr}[\rho\,\sigma_{j_1}\otimes\sigma_{j_2}].$ \\
On applying suitable local unitary operations, the correlation tensor becomes diagonalized:
 \begin{equation}\label{st41}
\small{\varrho}^{'}=\small{\frac{1}{4}(\mathbb{I}_{2\times2}+\vec{\mathfrak{a}}.\vec{\sigma}\otimes \mathbb{I}_2+\mathbb{I}_2\otimes \vec{\mathfrak{b}}.\vec{\sigma}+\sum_{j=1}^{3}\mathfrak{t}_{jj}\sigma_{j}\otimes\sigma_{j})},
\end{equation}
Here the correlation tensor is $T$$=$$\textmd{diag}(t_{11},t_{22},t_{33}).$
Under local unitary operations entanglement content of a quantum state remains invariant. Hence, steerability of $\varrho$ and $\varrho^{'}$ remain the same.
\subsection{Steering Inequality}
A linear steering inequality was derived in \cite{red3}. Under the
assumption that both the parties sharing a bipartite state($\rho_{AB}$) perform $n$ dichotomic quantum
measurements(on their respective particles), Cavalcanti, Jones, Wiseman, and Reid(CJWR) formulated a
series of correlators based inequalities\cite{red3} for checking steerability of $\rho_{AB}:$
\begin{equation}\label{mon2}
\mathcal{F}_{n}(\rho_{AB},\nu) = \frac{1}{\sqrt{n}}|\sum_{l=1}^{n} \langle A_{l} \otimes B_{l} \rangle | \leq 1
\end{equation}
Notations used in the above inequality are detailed below:
\begin{itemize}
  \item $\langle A_{l} \otimes B_{l} \rangle = \textmd{Tr}(\rho_{AB} (A_{l} \otimes B_{l}))$
  \item $\rho_{AB} \in \mathbb{H_{A}} \otimes \mathbb{H_{B}}$  is any bipartite quantum state\cite{costa}.
  \item  $A_{l} $$=$$ \hat{a}_{l}\cdot \overrightarrow{\sigma}$, $B_{l}$$=$$\hat{b}_{l} \cdot \overrightarrow{\sigma}$, $\hat{a}_{l},\,\hat{b}_{l} \in \mathbb{R}^{3}$  denote real orthonormal vectors. $A_{l} B_{l}$ thus denote inputs of Alice and Bob.
  \item $\nu =\{\hat{a}_{1},\hat{a}_{2},....\hat{a}_{n}, \hat{b}_{1},\hat{b}_{2},...,\hat{b}_{n} \}$ stands for the collection of measurement directions.
\end{itemize}
In case, dimension of each of local Hilbert spaces $\mathbb{H_{A}},\mathbb{H_{B}}$ is $2,$ $\rho_{AB}$ is given by Eq.(\ref{st4}). Violation of Eq.(\ref{mon2}) guarantees both way steerability of $\rho_{AB}$ in the sense that it is steerable from A to B and vice versa.  \\
Steering phenomenon remaining invariant under local unitary transformations, the analytical expressions of the
steering inequality remain unaltered if the simplified form(Eq.(\ref{st41})) of two qubit state
$\rho_{AB}$ is considered. The analytical expression of the upper bound of corresponding inequality for $3$ settings is given by\cite{costa}:
\begin{eqnarray}\label{string}
\textmd{Max}_{\nu}\mathcal{ F}_{3}(\rho_{AB},\nu)&=& \sqrt{\mathfrak{t}_{11}^{2}+ \mathfrak{t}_{22}^{2}+\mathfrak{t}_{33}^2},\nonumber\\
&=&\sqrt{\textmd{Tr}(T^tT)}\nonumber\\
&=&\sqrt{\textmd{Tr}(W^tW)}
\end{eqnarray}
where $W$ and $T$  denote the correlation tensor corresponding to density matrix representation of $\rho_{AB}$ given by Eq.(\ref{st4}) and Eq.(\ref{st41}) respectively. Last equality in Eq.(\ref{string}) holds as trace of a matrix is unitary equivalent. Hence, by the linear inequality(Eq.\ref{mon2}) (for $n$$=$$3$), any two qubit state $\rho_{AB}$(shared between $A$ and $B$) is both-way $F_{3}$ steerable if:
\begin{equation}\label{mon8}
\mathcal{S}_{AB} = Tr[T^{T}_{AB} T_{AB}] >1.
\end{equation}
Eq.(\ref{mon8}) gives only a sufficient condition detecting steerability. So if any state violates Eq.(\ref{mon8}), the state may be steerable, but its steerability remains undetected by CJWR inequality(Eq.(\ref{mon2}) for $n$$=$$3$). Any state violating Eq.(\ref{mon8}) may be referred to as $F_3$ unsteerable state in the sense that the state is unsteerable up to CJWR inequality for three settings.
\subsection{Bell Nonlocality in Three Settings Measurement Scenario}
Consider a bipartite measurement scenario involving three dichotomic measurements
settings(on each side). Such a scenario is often referred to as $(3,3,2,2)$ measurement scenario.
CHSH is not the only possible facet inequality in $(3,3,2,2)$
scenario\cite{garge1,garge2}. A complete list of facet inequalities of Bell polytope(for this measurement scenario)
was computed in \cite{garge2}. There exists only one Bell inequality inequivalent to CHSH inequality. That inequivalent facet inequality is referred to as the $I_{3322}$ inequality\cite{i3322}. Denoting local measurements of Alice and
 Bob as $A_1,A_2,A_3$ and $B_1,B_2,B_3$ respectively and the outcomes of each of this measurement
 as $\pm 1,$ $I_{3322}$ inequality takes the form\cite{i3322}:
\begin{eqnarray}\label{i32}
\small{-2P_{B_1}-P_{B_2}-P_{A_1}+P_{A_1B_1}+P_{A_1B_2}}&+&\small{P_{A_1B_3}+P_{A_2B_1}+}\nonumber\\
\small{P_{A_2B_2}-P_{A_2B_3}+P_{A_3B_1}-P_{A_3B_2}}&\leq&0,
\end{eqnarray}
where $\forall\, i,j$$=$$1,2,3,$ $P_{B_i}$$=$$p(1|B_i),$ $P_{A_i}$$=$$p(1|A_i)$ denote marginal probabilities and $P_{A_iB_j}$$=$$p(11|A_iB_j)$ stands for the joint probability terms. In terms of these probability terms, CHSH inequality\cite{cle} takes the form:
\begin{equation}\label{chsh}
   -(P_{A_1}+P_{B_1}+P_{A_2B_2})+P_{A_1B_1}+P_{A_1B_2}+P_{A_2B_1}\leq 0
\end{equation}
There exist quantum states which violate above inequality(Eq.(\ref{i32})) but satisfy
CHSH inequality(Eq.(\ref{chsh})) and vice-versa\cite{i3322}. Violation of anyone of
CHSH(Eq.(\ref{chsh})) or $I_{3322}$ inequality(Eq.(\ref{i32})) guarantees nonlocality of
corresponding correlations in $(3,3,2,2)$ scenario. Conversely, as these two are the only
inequivalent facet inequalities of
Bell-local polytope, so any correlation satisfying both Eqs.(\ref{i32},\ref{chsh}) is
Bell-local in $(3,3,2,2)$ scenario.
\subsection{Reduced Steering}
Notion of reduced steering has emerged in context of manifesting multipartite steering with the help of bipartite steering\cite{revw16}. Consider an $n$-partite quantum state $\varrho_{1,2,...,n}$ shared between $n$ parties $A_1,A_2,...,A_n$ If any one of these parties $A_i$(say) can steer the particle of another party say $A_j(i$$\neq$$ j)$ without aid of any of the remaining parties $A_k(k$$\neq$$i,j$), then the $n$-partite original state $\varrho_{1,2,...,n}$ is said to exhibit reduced steering. So reduced steering is one notion of steerability of $\varrho_{1,2,...,n}.$ Technically speaking $\varrho_{1,2,...,n}$ is steerable if at least one of the bipartite reduced states $\varrho_{i,j}$ is steerable.
\section{Hidden Steerability in Linear Network}\label{hid}
As already mentioned before, we focus on steering activation in quantum network scenario involving qubits such that steerable correlations are generated between two distant parties who do not share any direct common past. We start with an entanglement swapping network involving three parties.  \\
\subsection{Linear Three Party Network Scenario}
\begin{figure}[htb]
\centering
\includegraphics[width=3in]{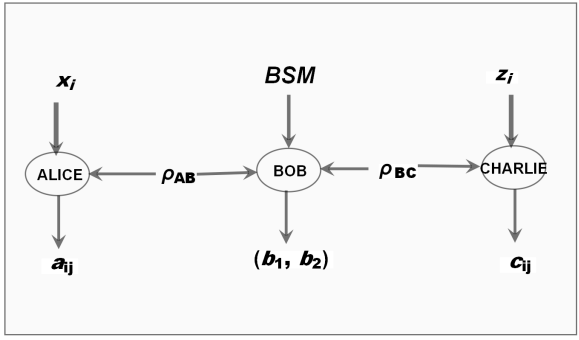}\\
\caption{\emph{A network of three parties Alice, Bob and Charlie. Alice and Bob share an entangled state $\rho_{AB}$ and that the state shared between Bob and Charlie is $\rho_{BC}$. Bob performs Bell basis measurement($BSM$) on his two particles and communicates the results to Alice and Charlie who then perform projective measurements on their conditional state.}}
\label{fig1}
\end{figure}
Consider a network of three parties Alice, Bob and Charlie arranged in a linear chain(see Fig.\ref{fig1}). Let $\rho_{AB}$ denote the entangled state shared between Alice and Bob whereas entangled state $\rho_{BC}$ be shared between Bob and Charlie. So initially Alice and Charlie do not share any physical state. Let one way classical communication be allowed between parties sharing a state. To be more specific Bob can communicate to each of Alice and Charlie. Alice and Charlie are thus the two non interacting parties.\\
 First Bob performs joint measurement on his two qubits in the Bell basis:
\begin{equation}\label{bell}
     |\phi^{\pm}\rangle =\frac{|00\rangle\pm|11\rangle}{\sqrt{2}},\,|\psi^{\pm}\rangle = \frac{|01\rangle\pm|10\rangle}{\sqrt{2}}.
 \end{equation}
Let $\vec{v}$$=$$(b_1b_2)$ denote the outcome of Bob: $(0,0),(0,1),(1,0),(1,1)$ correspond
to $ |\phi^{+}\rangle,$ $ |\phi^{-}\rangle,$ $ |\psi^{+}\rangle$ and $|\psi^{-}\rangle.$
Bob then communicates the results to Alice and Charlie. Let $\rho_{AC}^{(b_1b_2)}$ be the conditional state shared between Alice and Charlie when Bob obtains the outcome $\vec{b}$$=$$(b_1b_2).$ Each of Alice and Charlie now performs one of three arbitrary projective measurements on their respective qubits. Let $x_i$ and $z_i(i$$=$$1,2,3)$ denote the measurement settings of Alice and Charlie with $a_{ij}$ and $c_{ij}(j$$=$$0,1)$ denoting the binary outputs corresponding to $x_i$ and $z_j$ respectively. Bipartite correlations arising from the local measurements of Alice and Charlie are then used to test CJWR inequality for three settings:
\begin{equation}\label{mon21}
 \frac{1}{\sqrt{3}}|\langle A_{1} \otimes C_{1} \rangle+\langle A_{2} \otimes C_{2} \rangle+\langle A_{3} \otimes C_{3} \rangle | \leq 1
\end{equation} \\
Such a testing of the conditional states is required to check activation of steerability in the network. Idea of steerability activation detection is detailed below.
\subsection{Steering Activation in Network}
Phenomenon of steering activation is observed if both the initial states $\rho_{AB}$ and $\rho_{BC}$ are $F_3$ unsteerable whereas at least one of the four conditional states $\rho_{AC}^{(00)},\rho_{AC}^{(01)},\rho_{AC}^{(10)},\rho_{AC}^{(11)}$ is $F_3$ steerable. Precisely speaking, activation occurs if both $\rho_{AB}$ and $\rho_{BC}$ violate Eq.(\ref{mon8}) whereas $\rho_{AC}^{b_1b_2}$ satisfies the same for at least one possible pair $(b_1,b_2).$
Any pure entangled state being $F_3$ steerable, no activation is possible if one or both of the initial states $\rho_{AB}$ and $\rho_{BC}$ possess pure entanglement. So the periphery of analyzing steerability activation encompasses only mixed entangled states. We next provide with an instance of activation observed in the network.
\subsection{An Instance of Activation}\label{ins}
\begin{figure}[htb]
\centering
\includegraphics[width=3in]{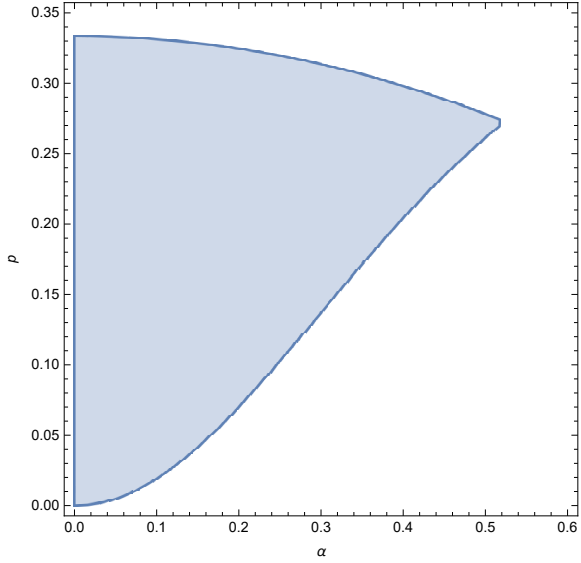}\\
\caption{\emph{Shaded region is a subspace in the parameter space $(p,\alpha)$ of the family of
states given by Eqs.(\ref{s14},\ref{sz14}). It indicates region of steering activation(as detected by
Eq.(\ref{mon8})) obtained in the entanglement swapping protocol(Fig.\ref{fig1}) when Bob obtains either $|\phi^{+}\rangle$ or $|\phi^{-}\rangle.$ It should be noted here that none of the conditional states $\rho_{AC}^{(00)},$ $\rho_{AC}^{(01)}$ is Bell nonlocal in three binary measurement settings scenario\cite{i3322}.}}
\label{fig2}
\end{figure}

Let us now consider the following families of two qubit states:
 \begin{equation}\label{s14}
    \gamma_1=(1-p)|\varphi\rangle\langle\varphi|+p|00\rangle\langle00|
 \end{equation}
 \begin{equation}\label{sz14}
\gamma_2=(1-p)|\varphi\rangle\langle\varphi|+p|11\rangle\langle11|
\end{equation}
where $|\varphi\rangle=\sin\alpha|01\rangle+\cos\alpha|10\rangle,$ $0\leq \alpha\leq\frac{\pi}{4}$ and $0\leq p\leq1$. These class of states were used for the purpose of increasing maximally entangled fraction in an entanglement swapping network\cite{mod}. Each of these families violates Eq.(\ref{mon8}) if:
\begin{equation}\label{s141}
    2 ((1-p) \sin 2\alpha)^2 + (2 p-1)^2\leq 1
 \end{equation}
 Now let $\rho_{AB}$ and $\rho_{BC}$ be any member of the family given by $\gamma_1$ and
 $\gamma_2$(Eqs.(\ref{s14},\ref{sz14})) respectively such that the state parameters satisfy Eq.(\ref{s141}). When Bob's particles get projected along $|\phi^{\pm}\rangle,$ each of the conditional states $\rho_{AC}^{00},\rho_{AC}^{01}$ is steerable(satisfying Eq.(\ref{mon8})) if:
 \begin{eqnarray}\label{s142}
    \frac{1}{N_1}(\small{9 - 26 p + 25 p^2+ 4 (3 - 8 p + 5 p^2) \cos(2\alpha)}\nonumber\\
    \small{  +
 3 (-1 + p)^2 \cos(4\alpha))}>1
 \end{eqnarray}
 where $N_1$$=$$2 (-1 - p + (-1 + p) \cos(2\alpha))^2.$
  Similarly if Bob's output is $|\psi^{\pm}\rangle,$ steerability of each of $\rho_{AC}^{10},\rho_{AC}^{11}$ is guaranteed if
\begin{equation}\label{s143}
    \frac{1}{N_2}(8 (-1 + p)^4 \sin(2 \alpha)^4+N_3)>1,
\end{equation}
 where $\small{N_2}$$=$$\small{(3 - 2 p + 3 p^2 - 4 (-1 + p) p \cos(2 \alpha) +}$\\
  $\small{(-1 + p)^2 \cos(4 \alpha))^2}$ and $N_3$$=$$ (3 - 10 p + 11 p^2 + 4 (-1 + p) p \cos(2\alpha) + (-1 + p)^2 \cos(4\alpha))^2.$
There exist state parameters $(p,\alpha)$ which satisfy Eqs.(\ref{s141},\ref{s142}). This in
turn indicates that there exist states from the two families(Eqs.(\ref{s14},\ref{sz14})) for which
steerability is activated for Bob obtaining $00$ or $01$ output(see Fig.\ref{fig2}). For example, activation
is observed for all members from these two families characterized by
$\alpha$$=$$0.1,$ and $p$$\in$$(0.001,0.331).$ However, in case conditional state $\rho_{AC}^{(10)}$ or $\rho_{AC}^{(11)}$ is obtained, activation of steering is not observed.\\
To this end one may note that a conditional state satisfying anyone of Eq.(\ref{s142}) or Eq.(\ref{s143}) is
Bell-local in $(3,3,2,2,)$ scenario, i.e., it violates neither $I_{3322}$ inequality(Eq.(\ref{i32}))
nor CHSH inequality(Eq.(\ref{chsh})).
\subsection{Measurement Settings Detecting Steerability}\label{mes}
As already mentioned before, for the purpose of investigating activation, criterion(Eq.(\ref{mon8}))
used as a sufficient criterion for detecting steerability of conditional states is a closed form of the upper bound of violation of CJWR inequality  for three settings(Eq.(\ref{mon21})).
It may be pointed out that the two parties sharing the conditional state in the network
being Alice and Charlie, in Eq.(\ref{mon21}), observables $C_i$ considered unlike that of $B_i$(used in Eq.(\ref{mon2})).
Now, as the closed form involves only state parameters \cite{costa}, in case any state satisfies the criterion given by Eq.(\ref{mon8}), state is steerable. But no information about measurement settings involved in detecting steerability of the state can be obtained.
However, from practical view point, it is interesting to know suitable measurement settings which help in steering the states. For that, given a two qubit state, suitable measurement settings are those projective measurements(for each of the two parties) for which the state considered violates
Eq.(\ref{mon21})$.\,A_i$$=$$\vec{a}_i.\vec{\sigma}$ and $C_i$$=$$\vec{c}_i.\vec{\sigma}$($i$$=$$1,2,3$)
denote projection measurements of Alice and Charlie respectively. As mentioned in section(\ref{pre1}), for
violation of CJWR inequality(Eq.(\ref{mon21})), each of Alice and Charlie performs projective measurements
in orthogonal directions: $\vec{a_i}.\vec{a_j}$$=$$0$$=$$\vec{c_i}.\vec{c_j},\,\forall\,i$$\neq$$ j.$ CJWR
inequality being symmetric\cite{red3}, violation of Eq.(\ref{mon21}) implies that the corresponding state
is steerable from Alice to Charlie and also from Charlie to Alice. Now, for obvious reasons choice of
appropriate settings is state specific. For providing some specific examples of suitable measurement
settings, we next consider the instance of activation provided in subsection(\ref{ins}).\\
Consider a particular member from each of the two families(Eqs.(\ref{s14},\ref{sz14})) characterized by
$(p,\alpha)$$=$$(0.214,0.267)$. None of these two states is steerable(up to Eq.(\ref{mon8})). So none of
these two states violate Eq.(\ref{mon21}). Let these two states be used in the linear network. In case
Bob gets output $(0,0)$ or $(0,1)$, conditional state $\rho_{AC}^{00}$ or $\rho_{AC}^{01},$ shared between
Alice and Charlie, violates Eq.(\ref{mon21}) when Alice projects her particle in anyone of the three
following orthogonal directions: \\
$(0,0,1),$ $(0,-1,0),$ $(-1,0,0)$ and
Charlie's projective measurement directions are given by:\\
 $(0,0,1),$ $(0,1,0),$ and
 $(-1,0,0).$ It may be noted here that these are not the only possible directions for which violation of Eq.(\ref{mon21}) is observed. Alternate measurement directions may also exist. However, there exists no measurement settings of Alice and Charlie for which the conditional states $\rho_{AC}^{10}$ or $\rho_{AC}^{11}$ violate Eq.(\ref{mon21}). So steering activation is possible(up to Eq.(\ref{mon21})) in case Bob obtains output $00$ or $01$ only.\\
Getting instances of steering activation in the network, an obvious question arises next:
can hidden steerability be observed for arbitrary two qubit states? This however turns out to be impossible in three measurement setting projective measurement scenario(for the non interacting parties) when one uses Eq.(\ref{mon8}) as steerability detection criterion\cite{costa}. We now analyze arbitrary two qubit states in this context.
\subsection{Characterization of Arbitrary Two Qubit States }\label{char}
Let two arbitrary states be initially considered in the swapping protocol. In density matrix formalism
the states are represented as:
 \begin{equation}\label{ss1441}
    \rho_{AB}=\small{\frac{1}{4}(\mathbb{I}_{2\times2}+\vec{\mathfrak{u}_1}.\vec{\sigma}\otimes \mathbb{I}_2+\mathbb{I}_2\otimes \vec{\mathfrak{v}_1}.\vec{\sigma}+\sum_{j_1,j_2=1}^{3}w_{1j_1j_2}\sigma_{j_1}\otimes\sigma_{j_2})},
\end{equation}
 \begin{equation}\label{ss1442}
    \rho_{BC}=\small{\frac{1}{4}(\mathbb{I}_{2\times2}+\vec{\mathfrak{u}_2}.\vec{\sigma}\otimes \mathbb{I}_2+\mathbb{I}_2\otimes \vec{\mathfrak{v}_2}.\vec{\sigma}+\sum_{j_1,j_2=1}^{3}w_{2j_1j_2}\sigma_{j_1}\otimes\sigma_{j_2})},
\end{equation}
Steerability of states remain unhindered under local unitary operations. Let suitable local unitary operations be applied to the initial states for diagonalizing the correlation tensors:
 \begin{equation}\label{s1441}
    \rho_{AB}^{'}=\small{\frac{1}{4}(\mathbb{I}_{2\times2}+\vec{\mathfrak{a}_1}.\vec{\sigma}\otimes \mathbb{I}_2+\mathbb{I}_2\otimes \vec{\mathfrak{b}_1}.\vec{\sigma}+\sum_{j=1}^{3}\mathfrak{t}_{1jj}\sigma_{j}\otimes\sigma_{j})},
\end{equation}
 \begin{equation}\label{s1442}
    \rho_{BC}^{'}=\small{\frac{1}{4}(\mathbb{I}_{2\times2}+\vec{\mathfrak{a}_2}.\vec{\sigma}\otimes \mathbb{I}_2+\mathbb{I}_2\otimes \vec{\mathfrak{b}_2}.\vec{\sigma}+\sum_{j=1}^{3}\mathfrak{t}_{2jj}\sigma_{j}\otimes\sigma_{j})},
\end{equation}
Let both $\rho_{AB}^{'}$ and $\rho_{BC}^{'}$ be $F_3$ unsteerable, i.e., let both of them violate  Eq.(\ref{mon8}). Hence  $\sum_{j=1}^3\small{\sqrt{\mathfrak{t}^2_{1jj}}}$$\small{\leq}$$1$ ,$\small{\sqrt{\mathfrak{t}^2_{2jj}}}$$\small{\leq}$$1.$ We next characterize $\rho_{AB}^{'}$ and $\rho_{BC}^{'}$ by analyzing nature of the conditional states $\rho_{AC}^{b_1b_2}.$ In this context, we provide three results each of which can be considered as a condition for no steering activation in the network. To be precise, if bloch parameters of any initial two qubit states satisfy assumptions(see Table \ref{table:ta20} for more details) of any of these three results then there will be no activation of $F_3$ steerability. Of these three results, two are proved analytically whereas the last one is a numerical observation only. First we give the two analytic results in form of two theorems.\\
\textbf{Theorem.1:}\textit{ If one or both the initial states(Eqs.(\ref{s1441},\ref{s1442})) do not have any non null local bloch vector(see Table \ref{table:ta20}) then none of the conditional states $\rho_{AC}^{b_1b_2}$ satisfies Eq.(\ref{mon8}).}\\
\textit{Proof:}See Appendix.A\\
Up to the steering criterion given by Eq.(\ref{mon8}), above result implies impossibility of steering activation in swapping network involving two qubit states whose local bloch vectors(corresponding to both the parties) vanish under suitable local unitary operations$.\,$Maximally mixed marginals class of two qubit states has no local bloch vector. So, activation is not possible in network involving any member from this class. \\
So hidden steerability cannot be exploited in absence of local bloch vectors corresponding to both the parties of a bipartite quantum state. But can the same be generated if both $\rho_{AB}^{'}$ and $\rho_{BC}^{'}$ has one non null local bloch vector? Following theorem provides a negative observation.\\
\textbf{Theorem.2:} \textit{If both the initial states  $ \rho_{AB}^{'}$ and $\rho_{BC}^{'}$ have only one non null local bloch vector, i.e., $\vec{\mathfrak{a}_1}$$=$$\vec{\mathfrak{a}_2}$$=$$\Theta$ or $\vec{\mathfrak{b}_1}$$=$$\vec{\mathfrak{b}_2}$$=$$\Theta$($\Theta$ denote null vector) then none of the conditional states $\rho_{AC}^{b_1b_2}$ satisfies Eq.(\ref{mon8}).} \\
\textit{Proof of Theorem.2:} This proof is exactly the same as that for Theorem.1 owing to the fact that here also   $\sqrt{\textmd{Tr}(\mathcal{V}_{b_1b_2}^T\mathcal{V}_{b_1b_2})}$$=$$\sqrt{\sum_{k=1}^3(t_{1kk}t_{2kk})^2}$ where $\mathcal{V}_{b_1b_2}$ denote correlation tensor of resulting conditional states $\rho_{AC}^{(b_1b_2)}$.\\
Note that in Theorem.2, $\vec{\mathfrak{a}_1}$$=$$\vec{\mathfrak{a}_2}$$=$$\Theta$ or $\vec{\mathfrak{b}_1}$$=$$\vec{\mathfrak{b}_2}$$=$$\Theta$ is considered. But what if $\vec{\mathfrak{a}_1}$$=$$\vec{\mathfrak{b}_2}$$=$$\Theta$ or $\vec{\mathfrak{b}_1}$$=$$\vec{\mathfrak{a}_2}$$=$$\Theta?$ Does activation occurs in such case?
Numerical evidence suggests a negative response to this query:\\
\textbf{Numerical Observation:} \textit{If $\vec{\mathfrak{a}_1}$$=$$\vec{\mathfrak{b}_2}$$=$$\Theta$ or $\vec{\mathfrak{b}_1}$$=$$\vec{\mathfrak{a}_2}$$=$$\Theta$ then none of the conditional states $\rho_{AC}^{b_1b_2}$ satisfies Eq.(\ref{mon8}).}\\
Justification of this observation is based on the fact that numerical maximization of the steerability expression(Eq.(\ref{mon8})) corresponding to each possible conditional state $\rho_{AC}^{b_1b_2}$ gives $1$ under the constraints that both the initial quantum states($\rho_{AB}^{'},\rho_{BC}^{'}$). Consequently none of the conditional states satisfies Eq.(\ref{mon8}) if none of $\rho_{AB}^{'},\rho_{BC}^{'}$ satisfies Eq.(\ref{mon8}). \\
Above analysis points out the fact that for revealing hidden steerability, each of $\rho_{AB}^{'}$ and $\rho_{BC}^{'}$ should have non null local bloch vectors corresponding to both the parties. However that condition is also not sufficient for activation. In case, correlation tensor of any one of them is a null matrix, the state is a separable state. When such a state is considered as an initial state in the network, none of the conditional states is entangled and thereby activation of steerability becomes impossible. So, when steerability is activated in the network following are the necessary requirements:  \\
\begin{itemize}
  \item All of the local bloch vectors must be non null: $\vec{\mathfrak{a}_i}$$\neq$$\Theta$,$\vec{\mathfrak{b}_i}$$\neq$$\Theta\,\forall\,i$ and
  \item Both the initial states should have non null correlation tensors.
 \end{itemize}
However the above conditions are only necessary for activation purpose but are not sufficient for the same. We next provide illustration with specific examples in support of our claim.\\
\begin{center}
\begin{table}[htp]
\caption{Assumptions of three results(analyzed above) are enlisted here. The correlation tensor of each of the two initial states $\rho_{AB}^{'}$ and $\rho_{BC}^{'}$ remain arbitrary. Restrictions are imposed over the local bloch parameters only. }
\begin{center}
\begin{tabular}{|c|c|c|}
\hline
Result&Assumptions&Steerability\\
&&Activation\\
\hline
&$(\vec{\mathfrak{a}_i},\vec{\mathfrak{b}_i})$$=$$(\Theta,\Theta)\,\forall\,i$&\\
&\small{or}&\\
&$(\vec{\mathfrak{a}_i},\vec{\mathfrak{b}_i})$$=$$(\Theta,\Theta)$ for $i$$=$$1$ &\\
\small{Theorem.1}&\small{or}&No\\
&$(\vec{\mathfrak{a}_i},\vec{\mathfrak{b}_i})$$=$$(\Theta,\Theta)$ for $i$$=$$2$ &\\
\hline
&$\vec{\mathfrak{a}_1}$$=$$\vec{\mathfrak{a}_2}$$=$$\Theta$&\\
\small{Theorem.2}&\small{or}&No\\
&$\vec{\mathfrak{b}_1}$$=$$\vec{\mathfrak{b}_2}$$=$$\Theta$ &\\
\hline
 &$\vec{\mathfrak{a}_1}$$=$$\vec{\mathfrak{b}_2}$$=$$\Theta$ &\\
\small{Numerical}&\small{or}&\\
\small{Observation}& $\vec{\mathfrak{b}_1}$$=$$\vec{\mathfrak{a}_2}$$=$$\Theta$&No\\
\hline
\end{tabular}
\end{center}
\label{table:ta20}
\end{table}
\end{center}
\subsubsection{\textbf{Illustration}}
Let us now analyze the classes of states given by Eqs.(\ref{s14},\ref{sz14}) in perspective of above characterization.
Both the families of initial states(Eqs.(\ref{s14},\ref{sz14})) have local bloch vectors: $\vec{\mathfrak{a}_1}$$=$$(0,0,p$-$\cos(2\alpha)(1$-$p)),$
$\vec{\mathfrak{b}_1}$$=$$(0,0,p$+$\cos(2\alpha)(1$-$p)),$ $\vec{\mathfrak{a}_2}$$=$$(0,0,$-$p$-$
\cos(2\alpha)(1$-$p)),
$ $\vec{\mathfrak{b}_2}$$=$$(0,0,$-$p$+$\cos(2\alpha)(1$-$p)).$ Local bloch vectors are non null for
$\cos(2\alpha)$$\neq$$\pm\frac{p}{1-p}.$ Correlation tensors of the states from both the families are given by $\textmd{diag}((1-p)\sin(2\alpha),(1-p)\sin(2\alpha),2p-1).$ Clearly activation is not observed for all family members having non null local blochs as well as non null correlation tensors. For instance, consider $(p,\alpha)$$=$$(0.6,0.6).$
Bloch parameters of corresponding states are given by:
\begin{itemize}
\item  $\vec{\mathfrak{a}_1}$$=$$(0,0,0.455057))$,$\vec{\mathfrak{b}_1}$$=$$(0,0,0.744943),$ \\
\item  $\vec{\mathfrak{a}_2}$$=$$(0,0,$-$0.455057))$, $\vec{\mathfrak{b}_2}$$=$$(0,0,$-$0.744943),$\\
\item $\textmd{diag}(t_{i11},t_{i22},t_{i33})$$=$$\textmd{diag}(0.372816,0.372816,0.2),\,\forall\, i$
\end{itemize}
 No steering activation is observed when these two states
are used in the network. This in turn implies that the
criteria given in \ref{char} are only necessary but not sufficient to ensure activation in the network. \\
Now, as already discussed in subsection(\ref{ins}),
there exist members from these families(see Fig.(\ref{fig2})) which when used in the swapping network
steering activation is observed. \\

Network scenario considered so far involved two states shared between three parties. However, will increasing length of the chain, hence increasing number of initial states be useful for the purpose of revealing hidden steerability? Though general response to this query is non trivial, we consider a star network configuration of four parties to give instances of activation of reduced steering.
\section{Non-Linear Swapping Network Involving $n$$\geq$$3$ States}\label{lin2}
\begin{figure}
\includegraphics[width=3.4in]{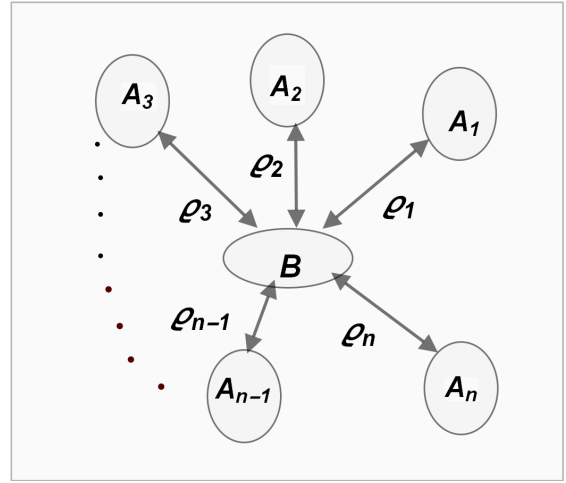}\\
\caption{\emph{Schematic Diagram of a star network. For $i$$=$$1,2,...,n,$ bipartite state $\varrho_i$ is
shared between parties $B$ and $A_i.$ Party $B$ performs joint measurement on state of his $n$
particles and communicates his output to each of $A_1,A_2,...,A_n.$ Reduced steering of corresponding
conditional state shared between $A_1,A_2,...,A_n$ is checked. }}
\label{starry1}
\end{figure}

\begin{center}
\begin{figure}
 \begin{tabular}{|c|c|}
 				\hline
 				\subfloat[Output: $\rho^{(1)}$]{\includegraphics[trim = 0mm 0mm 0mm 0mm,clip,scale=0.45]{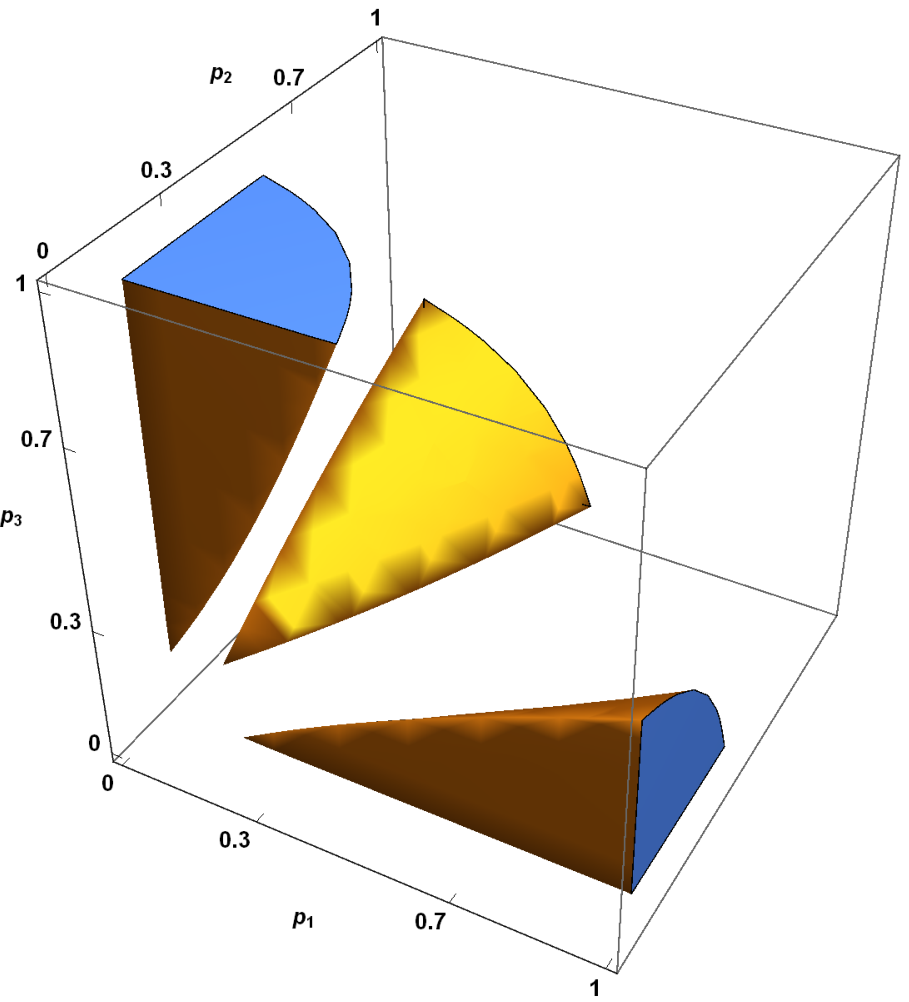}}
                &\subfloat[Output: $\rho^{(6)}$]{\includegraphics[trim = 0mm 0mm 0mm 0mm,clip,scale=0.45]{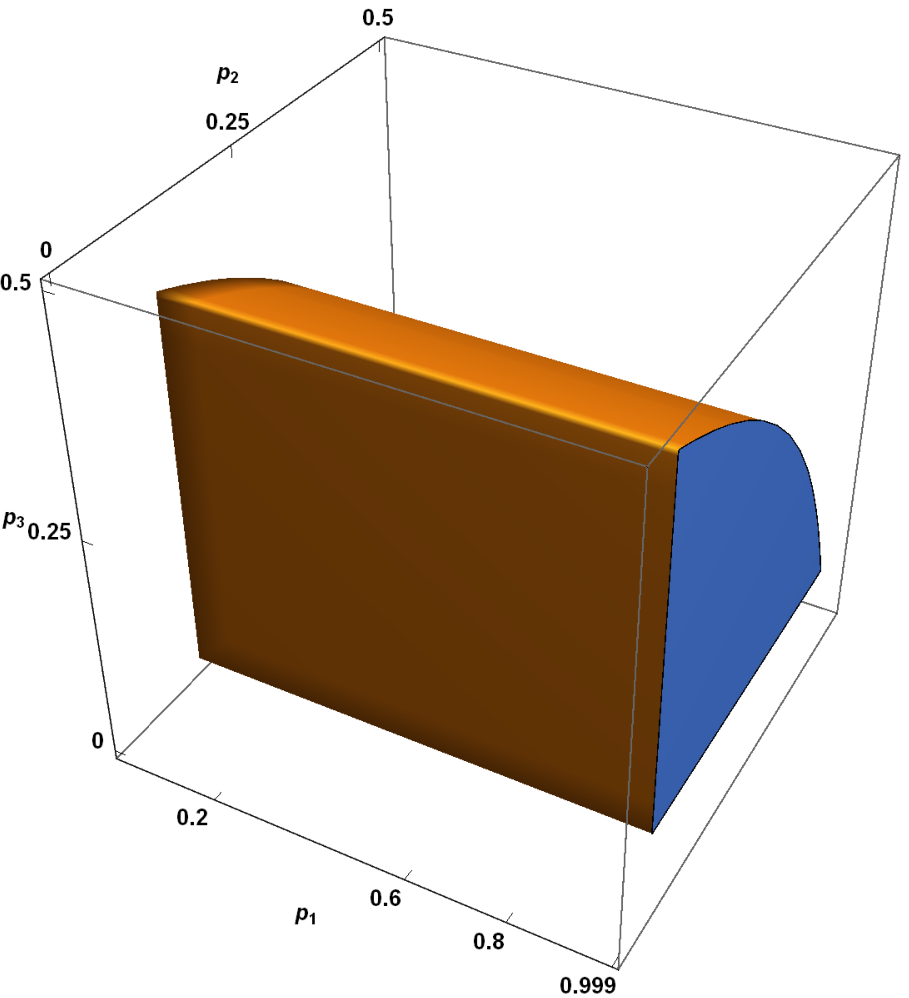}}\\
 				\hline
 				\subfloat[Output: $\rho^{(7)}$]{\includegraphics[trim = 0mm 0mm 0mm 0mm,clip,scale=0.45]{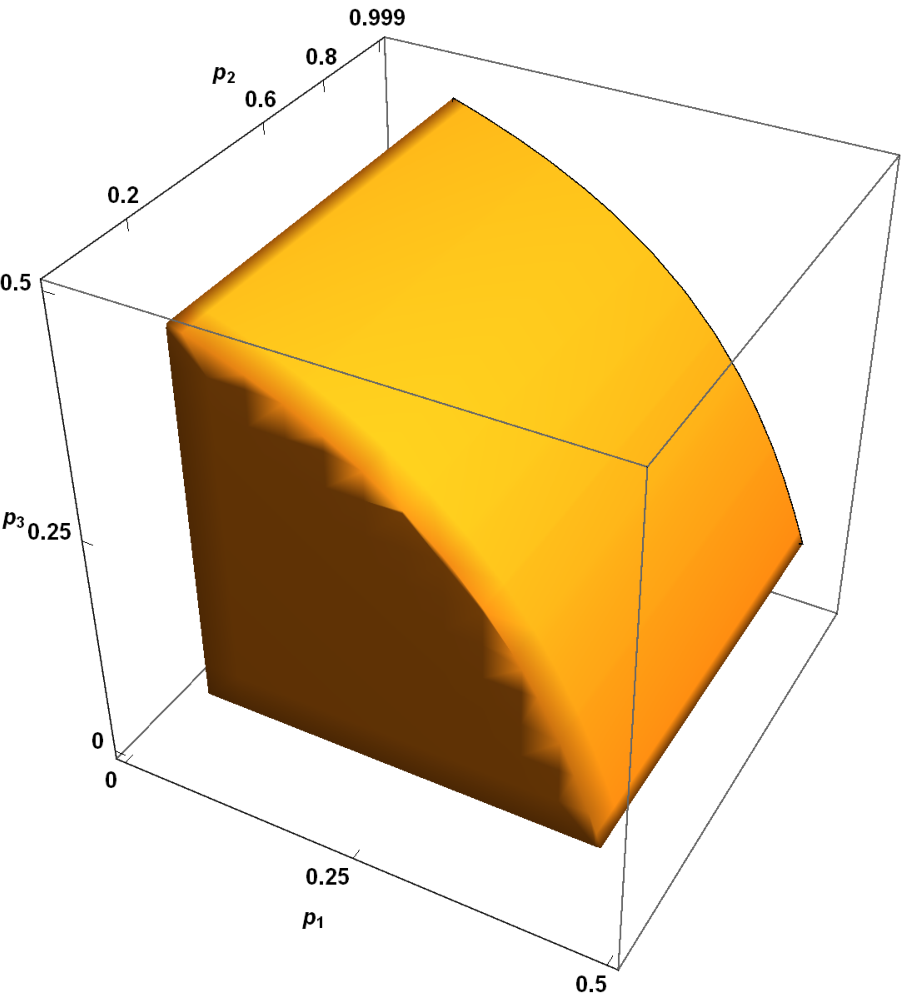}}
                &\subfloat[Output: $\rho^{(8)}$]{\includegraphics[trim = 0mm 0mm 0mm 0mm,clip,scale=0.45]{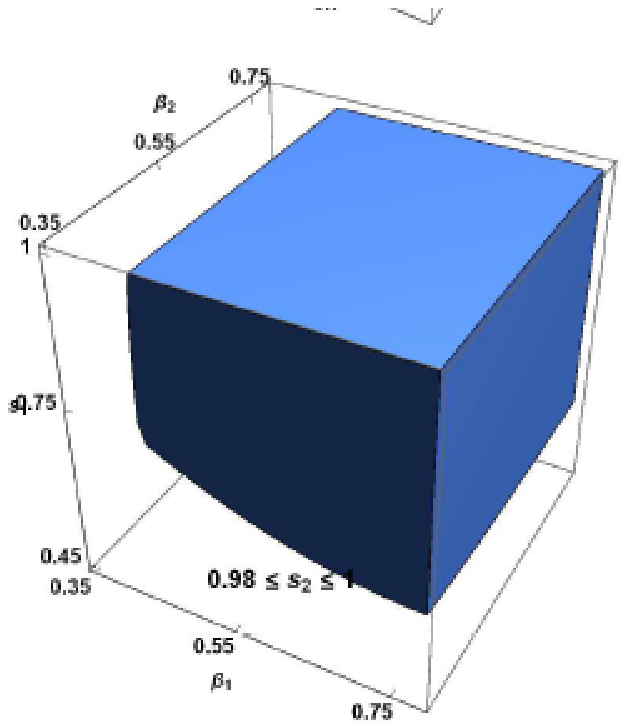}}\\
 				\hline
 \end{tabular}
 \caption{\emph{ Shaded region in each of four sub figures in the grid gives the steering activation region obtained stochastically depending on the different possible outputs of party $B$'s measurement in orthonormal basis(Eq.(\ref{basis2})). Here the star network scenario(Fig.\ref{starry1}) involves three non identical states from the class given by Eq.(\ref{s14}) for $\alpha$$=$$0.2.$ Starting from the top row and moving from left to right, shaded regions indicates
 reduced steering activation when $B$'s particles get projected along
 $\delta_1,\delta_6,\delta_7$ and $\delta_8$ respectively.}}
\label{starry2}
 \end{figure}
 \end{center}

Consider $n+1$($n$$\geq$$3$) number of parties $A_1,A_2,...,A_{n}$ and $B.$
Let $n$ bipartite states $\varrho_i(i$$=$$1,2,...,n)$ be shared
between the parties such that $\varrho_i$ is shared between parties $B$ and $A_i(i$$=$$1,2,...,n)$(see
Fig.\ref{starry1}). $B$ performs a joint measurement on his share of qubits from each $\varrho_i$ and
communicates outputs to the other parties $A_i(i$$=$$1,2,...,n).$ Reduced steering of each of the
conditional $n$-partite states is checked. To be precise, it is checked whether at least one possible
bipartite reduced state of at least one of the conditional states satisfies Eq.(\ref{mon8}). In case at
least one of the conditional states has reduced steering when none of $\varrho_{i}(i$$=$$1.2...,n)$ satisfies
Eq.(\ref{mon8}), activation of steerability is obtained. Activation is thus observed when one of the
$n$ parties sharing $n$-partite conditional state can steer the particles of another party without
any assistance from remaining $n-2$ parties sharing the same. \\
Consider a specific instance of $n$$=$$3.$ Let each of $\varrho_1,\varrho_2,\varrho_3$ be a member of the
family of states given by Eq.(\ref{s14}) with $p$$=$$p_1,p_2,p_3$ for $\varrho_1,\varrho_2,\varrho_3$
respectively. Let $B$ perform joint measurement in the  following orthonormal basis:
\begin{eqnarray}\label{basis2}
 |\delta_1\rangle =\frac{1}{\sqrt{3}}(|001\rangle+|100\rangle+|010\rangle)\nonumber\\
 |\delta_2\rangle = \frac{1}{\sqrt{3}}(|010\rangle-|100\rangle+|000\rangle)\nonumber\\
 |\delta_3\rangle =\frac{1}{\sqrt{3}}(-|010\rangle+|001\rangle+|000\rangle)\nonumber \\
 |\delta_4\rangle = \frac{1}{\sqrt{3}}(|100\rangle+|000\rangle-|001\rangle)\nonumber\\
 |\delta_5\rangle = \frac{1}{\sqrt{3}}(|101\rangle+|110\rangle+|011\rangle)\nonumber\\
  |\delta_6\rangle =\frac{1}{\sqrt{3}}(|110\rangle-|101\rangle+|111\rangle)\nonumber \\
  |\delta_7\rangle =\frac{1}{\sqrt{3}}(-|110\rangle+|111\rangle+|011\rangle)\nonumber\\
  |\delta_8\rangle = \frac{1}{\sqrt{3}}(|111\rangle+|101\rangle-|011\rangle)
\end{eqnarray}
When $B$'s particles get projected along $\delta_j,$ let $\rho^{(j)}(j$$=$$1,...,8)$ denote the conditional state shared between $A_1,A_2,A_3.$ Reduced steering of each of the conditional states is checked in terms of the steering inequality given by Eq.(\ref{mon8}). Now, let all three initial states $\varrho_1,\varrho_2,\varrho_3$ violate Eq.(\ref{mon8}). When $B$'s state gets projected along any one of $\delta_1,\delta_6,\delta_7,\delta_8$(Eq.(\ref{basis2})), for some state parameters $(p,\alpha),$ each of corresponding conditional states has reduced steering.
Region of activation is thus observed(see Fig.\ref{starry2}). Some particular instances of activation are enlisted in Table.(\ref{table:ta18}). At this point it should be pointed out that none of the reduced states corresponding to the conditional states violates neither $I_{3322}$ inequality(Eq.(\ref{i32})) nor CHSH inequality(Eq.(\ref{chsh})) and hence are Bell local(in $(3,3,2,2)$ scenario).

\begin{center}
\begin{table}[htp]
\caption{Some specific values of state parameters are enlisted here for which stochastic steering activation(in
terms of reduced steering) is observed in nonlinear network(Fig.\ref{starry1}). To be more precise, for
$\alpha$$=$$0.2,$ other parameters $p_1,p_2,p_3$ are specified for the three
 non identical states from the class given by Eq.(\ref{s14}). First column in the table gives the
  conditional state corresponding to which activation is observed.}
\begin{center}
\begin{tabular}{|c|c|c|c|}
\hline
State&$p_1$&$p_2$&Range\\
&&&of $p_3$\\
\hline
$\rho^{(1)}$&$0.08$&$0.075$&$(0.2,1]$\\
\hline
$\rho^{(6)}$&$0.08$&$0.075$&$(0.071,0.467]$\\
\hline
$\rho^{(7)}$&$0.08$&$0.075$&$(0,071,0.465]$\\
\hline
$\rho^{(8)}$&$0.08$&$0.075$&$(0.2,1)$\\
\hline
\end{tabular}
\end{center}
\label{table:ta18}
\end{table}
\end{center}

\section{genuine Activation of Steerability}\label{pur}
Most of the research works in the field of activation scenarios analyze activation of nonclassicality of quantum states with respect to any specific detection criterion of the nonclassical feature considered. To be precise, let $\mathcal{C}$ denote a detection criterion for a specific notion of quantum nonclassicality. Activation is said to be observed in any protocol if using one or more quantum states(or identical copies of the same state), none of which satisfies $\mathcal{C},$ another quantum state is generated(at the end of the protocol) that satisfies $\mathcal{C}.$ Using detection criterion of $\mathcal{F}_3$ steerability\cite{red3,costa}, so far we have obtained various cases of steering activation in both linear and nonlinear quantum networks. But quite obviously such a trend of activation analysis is criterion specific and in general can be referred to as \textit{activation of $\mathcal{F}_3$ steerability.} But here we approach to explore activation beyond the periphery of criterion specification. We refer to such activation as \textit{genuine activation of steerability}.\\
Let us consider the linear chain of three parties(Fig.\ref{fig1}). For genuine activation we use states
which satisfy some criterion of unsteerability and then explore $\mathcal{F}_3$ steerability of the
conditional states resulting due to Bell basis measurement(BSM) by the intermediate party(Bob) in the
protocol. Genuine activation of steerability occurs in case at least one of $\rho_{AC}^{b_1b_2}$ satisfies
Eq.(\ref{mon8}).
In \cite{unsteer}, the authors proposed an asymmetric sufficient criterion of bipartite unsteerability.\\
Let $\rho_{AB}$ be any two qubit state shared between Alice and Bob(say). In density matrix formalism
$\rho_{AB}$ is then provided by Eq.(\ref{st4}). Consider a positive, invertible linear map $\Lambda$,
whose action on $\rho_{AB}$ is given by\cite{unsteer}:
\begin{equation}\label{map1}
    \mathbb{I}_2\otimes \Lambda(\rho_{AB})=\mathbb{I}_2\otimes\rho_{B}^{-1}\rho_{AB}\mathbb{I}_2\otimes\rho_B^{-1},
\end{equation}
where $\mathbb{I}_2$ is $2\times2$ identity matrix in Hilbert space associated with $1^{st}$ party and
$\rho_B$$=$$\textmd{Tr}_{A}(\rho_{AB}).$ Let $\rho_{AB}^{(1)}$ denote the state density matrix obtained
after applying the above map to $\rho_{AB}.$ Local bloch vector corresponding to $2^{nd}$ party(Bob) of
$\rho_{AB}^{(1)}$ becomes a null vector\cite{unsteer}:
\begin{equation}\label{map2}
\small{\rho_{AB}^{(1)}}=\small{\frac{1}{4}(\mathbb{I}_{2\times2}+\vec{\mathfrak{u}^{'}}.\vec{\sigma}\otimes \mathbb{I}_2+\sum_{j_1,j_2=1}^{3}w_{j_1j_2}^{'}\sigma_{j_1}\otimes\sigma_{j_2})},
\end{equation}
On further application of local unitary operations to diagonalize correlation tensor, $\rho_{AB}^{'}$ ultimately becomes:
\begin{equation}\label{map3}
\small{\rho_{AB}^{(2)}}=\small{\frac{1}{4}(\mathbb{I}_{2\times2}+\vec{\mathfrak{u}^{''}}.\vec{\sigma}\otimes \mathbb{I}_2+\sum_{j=1}^{3}w_{jj}^{''}\sigma_{j}\otimes\sigma_{j})},
\end{equation}
$\rho_{AB}^{(2)}$(Eq.(\ref{map3})) is referred to as the \textit{canonical form} of $\rho_{AB}$ in \cite{unsteer} where the authors argued that $\rho_{AB}$ will be unsteerable if and only if $\rho_{AB}^{(2)}$ is unsteerable. They showed that $\rho_{AB}$ is unsteerable from Alice to Bob if\cite{unsteer}:
\begin{equation}\label{unstr}
    \textmd{Max}_{\hat{x}}((\vec{\mathfrak{a}}.\hat{x})^2+2 ||\mathcal{W}^{''}\hat{x}||)\leq 1
\end{equation}
where $\hat{x}$ is any unit vector indicating measurement direction, $\mathcal{W}^{''}$ denotes the
correlation tensor of $\rho_{AB}^{''}$ and $||.||$ denotes Euclidean norm.\\
For our purpose we consider the unsteerability criterion given by Eq.(\ref{unstr}). Below we characterize arbitrary two qubit states in ambit of genuine activation of steerability.
\subsection{Characterizing Two Qubit States}\label{char2}
Let $\rho_{AB}$ and $\rho_{BC}$(Eqs.(\ref{ss1441},\ref{ss1442})) denote two arbitrary two qubit states used in the network. It turns out that local bloch vector corresponding to first party of the initial states play a significant role in determining possibility of genuine activation of steering in the network. Next we give two results. While one of those is provided with an analytical proof, analysis of the other one relies on numerical optimization. We first state the analytical result.\\
\textbf{Theorem.3:} If canonical forms of both the initial states $\rho_{AB}$ and $\rho_{BC}$(Eqs.(\ref{ss1441},\ref{ss1442})) satisfy the unsteerability criterion(Eq.(\ref{unstr})),  then genuine activation of steerability is impossible if both of them have null local bloch vector corresponding to first party, i.e., $\vec{u_1},\vec{u_2}$$=$$\Theta.$ \\
\textit{Proof:} See appendix.\\
Genuine activation being impossible in case both $\vec{u_1},\vec{u_2}$ are null vectors, an obvious
question arises whether it is possible in case at least one of $\vec{u_1},\vec{u_2}$$\neq$$\Theta.$
We provide next result in this context. As numerical procedure is involved in corresponding calculations(see Appendix C), our next result will be considered as a numerical observation only.\\
\textbf{Numerical Observation:} \textit{If canonical forms of both $\rho_{AB}$ and $\rho_{BC}$(Eqs.(\ref{ss1441},\ref{ss1442})) satisfy the unsteerability criterion(Eq.(\ref{unstr})),  then genuine activation of steerability is impossible if any one of $\rho_{AB}$ or $\rho_{BC}$ has null local bloch vector corresponding to first party, i.e., at least one of $\vec{u_1},\vec{u_2}$$=$$\Theta.$ }\\
Justification of this observation is given in Appendix C\\
Clearly, the above two results, combined together provide a necessary criterion for genuine activation of steerability:\textit{When canonical forms of both the initial states satisfy Eq.(\ref{unstr}), if steering is genuinely activated in the network then both the initial states must have non null local bloch vectors corresponding to first party, i.e., $\vec{u_1}$$\neq$$\Theta,\,\vec{u_2}$$\neq$$\Theta.$} \\We next provide with examples in this context.
\subsection{Examples}
Consider a family of states\cite{unsteer}:
\begin{equation}\label{unstr1}
    \Omega= s|\chi\rangle\langle\chi|+(1-s)\Omega^{1}\otimes\frac{\mathbb{I}_{2}}{2},
\end{equation}
where $|\chi\rangle$$=$$\cos(\beta)|00\rangle+\sin(\beta)|11\rangle,$ $0$$\leq$$s$$\leq$$1,$ $\mathbb{I}_2$ is $2x2$ identity matrix in Hilbert space associated with $2^{nd}$ party and $\Omega^{1}$ is the reduced state of first party obtained by tracing out second party from $|\chi\rangle\langle \chi|,$ i.e., $\Omega^{1}$$=$$\cos^2(\beta)|0\rangle\langle 0|+\sin^2(\beta)|1\rangle\langle 1|.$\\
 For $\beta$$\neq$$\frac{\pi}{4},$ any member from this class has non null local bloch vector corresponding to first party:$(0,0,\cos(2\beta)).$ Canonical form(Eq.(\ref{map3})) of any member of this class satisfies Eq.(\ref{unstr}) if\cite{unsteer}:
\begin{equation}\label{unstr2}
    \cos^2(2\beta)\geq \frac{2s-1}{(2-s)s^3}.
\end{equation}
\\
\begin{figure}[htb]
\includegraphics[width=3.3in]{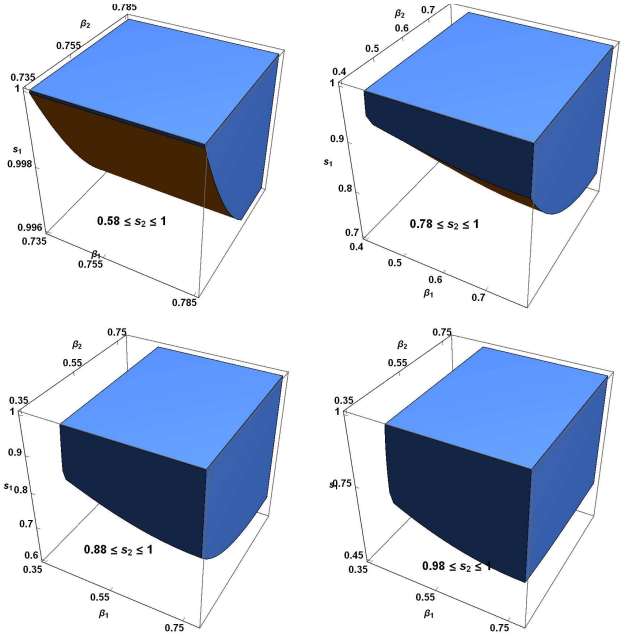}\\
\caption{\emph{Shaded regions in the sub figures give region of genuine activation of steerability for
different ranges of state parameter $s_2.$ None of the steerable conditional states obtained in the
protocol is Bell nonlocal in $(3,3,2,2)$ measurement scenario.}}
\label{pure1}
\end{figure}
Let two non identical members $\Omega_1$ and $\Omega_2$ from this class(Eq.(\ref{unstr1})) be used in
the entanglement swapping protocol(Fig.\ref{fig1}). Let $(\beta_1,s_1)$ and $(\beta_2,s_2)$ be state
parameters of $\Omega_1$ and $\Omega_2$ respectively. Let both $\Omega_1$ and $\Omega_2$ be unsteerable.
Now, for some values of the state parameters, the conditional states generated in the
protocol turn out to be steerable(see Fig.\ref{pure1}) as they satisfy Eq.(\ref{mon8}). Range of parameter $s_2$(for some fixed value of other three parameters $(\beta_1,\beta_2,s_1)$) for which  genuine
activation occurs, is provided in Table.\ref{table:ta19}. \\
\begin{center}
\begin{table}[htp]
\caption{ For some specific values of parameters $(\beta_1,\beta_2,s_1)$, of $\Omega_1,\Omega_2$
range of steerability activation other parameter $s_2$ is specified. First column specifies
the conditional state corresponding to which activation is observed. }
\begin{center}
\begin{tabular}{|c|c|c|c|c|}
\hline
State&$\beta_1$&$\beta_2$&$s_1$&Range \\
&&&&of $s_2$\\
\hline
$\rho^{00}_{AC}$&$0.75$&$0.76$&$0.99$&$[0.58,1]$\\
&&&&\\
\hline
$\rho^{01}_{AC}$&$0.65$&$0.6$&$0.97$&$[0.78,1]$\\
&&&&\\
\hline
$\rho^{10}_{AC}$&$0.55$&$0.55$&$0.9$&$[0.88,1]$\\
\hline
$\rho^{11}_{AC}$&$0.6$&$0.55$&$0.8$&$[0.98,1]$\\
&&&&\\
\hline
\end{tabular}
\end{center}
\label{table:ta19}
\end{table}
\end{center}
\begin{figure}
\includegraphics[width=3.1in]{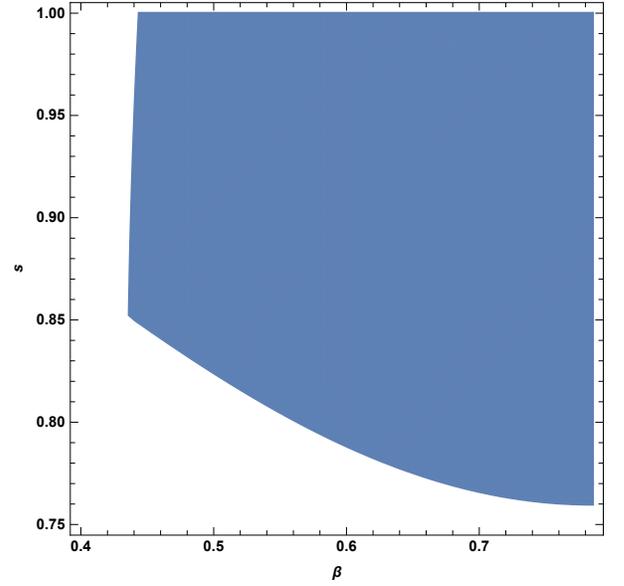}\\
\caption{\emph{Genuine activation region obtained for any possible conditional state when two identical copies of a state from $\Omega$ class are used in the network.}}
\label{pure2}
\end{figure}
Now, as discussed above, the criterion of both the initial unsteerable states having non null local bloch vector(corresponding to first party) is necessary for genuine activation. The criterion however turns out to be insufficient for the same. We next provide an example in support of our claim.\\
Consider two distinct members $\Omega_3,\Omega_4$ from the family of states given by Eq.(\ref{unstr1}) corresponding to the parameters: $(\beta_3,s_3)$$=$$(0.1,0.7)$ and $(\beta_4,s_4)$$=$$(0.3,0.59).$ Local bloch vectors(corresponding to first party) of $\Omega_3$ and $\Omega_4$ are $(0,0,0.980067)$
 and $(0,0,0.825336)$ respectively. Both of $\Omega_3$ and $\Omega_4$ satisfy the unsteerability criterion
 given by Eq.(\ref{unstr2}). These two states(in their canonical forms) are now used in the
 tripartite linear network. Bloch matrix representation of each of the conditional states are
 enlisted in Table.\ref{table:ta17}(see Appendix D). Unsteerability criterion(Eq.(\ref{unstr})) is then tested for each
 of these conditional states. The optimal value(obtained numerically) in the maximization problem involved in Eq.(\ref{unstr})
 turns out to be less than unity for each of the conditional states(Table.\ref{table:ta17}). Hence, all the conditional states are unsteerable. Consequently no genuine activation of steering is observed in the network using $\Omega_3$ and $\Omega_4.$ \\
It may be noted that genuine activation occurs for any possible output of Bob when two identical copies of
same state from this class are used in the network(see Fig.\ref{pure2}). For instance,
when two identical copies of $\Omega_1$ for $\beta_1$$=$$0.7$ are considered as initial states,
steerability is activated genuinely for $s_1$$\in$$(0.77,1].$
\section{Discussions}\label{conc}
In different information processing tasks, involving steerable correlations, better
efficiency of the related protocols(compared to their classical counterparts) basically rely upon
quantum entanglement. Though pure entanglement is the most suitable candidate, but owing to environmental effects, mixed entanglement is used in practical scenarios. In this context, any steerable mixed entangled state is considered to be useful. In case it fails to generate steerable correlations, it will be interesting to exploit its steerability(if any) by subjecting to suitable sequence of measurements. Entanglement swapping protocol turns out to be an useful tool in this perspective. Let us consider the two families of states given by Eqs.(\ref{s14},\ref{sz14}). Both of them are noisy versions of pure entangled states $\varphi.$ To be more specific, these families are obtained via amplitude damping of $\varphi$\cite{song}. As already discussed above, steering activation is obtained via entanglement swapping protocol for some members from these two families. This in turn point out that entanglement swapping protocol is useful in exploiting steerability from unsteerable(up to the steering criteria given by Eq.(\ref{mon8})) members from these two families. All such discussions in turn point out the utility of steerability activation in network scenarios from practical viewpoint. Characterization of arbitrary two qubits states will thus be helpful in exploiting utility of any given two qubit state in the ambit of steering activation(up to Eq.(\ref{mon8})). That steerability of depolarized noisy versions of pure entangled states cannot be activated(in approach considered here) is a direct consequence of such characterization owing to the fact that this class of noisy states has no local bloch vector. Apart from revealing hidden steerability, it will be interesting to explore whether the activation protocols can be implemented in any information processing task involving network scenario so as to render better results. \\
In \cite{cjwr3}, the authors have shown that if a two qubit state is $\mathcal{F}_3$ steerable, i.e., satisfies Eq.(\ref{mon8}), then it is useful for teleportation. This in turn points out the utility of the activation networks discussed here in perspective of information theoretic tasks. To be more precise, consider, for example, the tripartite linear network(Fig.\ref{fig1}). Both the initial states $\rho_{AB},\rho_{BC}$ used in the network violates Eq.(\ref{mon8}). So $\rho_{AB}$($\rho_{BC}$) cannot be used to teleport qubit from Alice to Bob(from Bob to Charlie). Now, if steerability is activated in the network stochastically, resulting conditional state can be used for the purpose of teleportation. In case, activation occurs for all possible outputs of Bob, any of the four conditional states turns out to be useful in teleportation protocol.\\
Now our analysis of activation in network scenarios is criterion specific and we have just provided partial characterization of two qubit state space in context of genuine activation of steerability. The unsteerability criterion\cite{unsteer} involves maximization over arbitrary measurement directions(Eq.(\ref{unstr})). Deriving closed form of this criterion, genuine activation of steerability can be analyzed further. In star network scenario choice of the specific orthonormal basis(Eq.(\ref{basis2})) for joint measurement by the central party($B$) served our purpose to show that increasing number of states non-linearly can yield better results compared to the standard three party network scenario(Fig.(\ref{fig1})). Also such activation scenario is significant as hidden steerability is revealed when at least one of the $n$ parties sharing $n$-partite conditional state can steer the particles of another party without co-operation from remaining $n-2$ parties. Further analysis of such form of steerability activation(via notion of reduced steering) using more general measurement settings of the central party $B$ will be a potential direction of future research. It will also be interesting to analyze a scheme of $m$ copies of bipartite states arranged in a linear chain where activation occurs only after projection on any of $n$$<$$m$ copies.\\
In \cite{new1}, the authors introduced notion of network steering  and network local hidden state(NLHS) models  in networks involving independent sources. They have provided with no-go results for network steering in a large class of network scenarios, by explicitly constructing NLHS models. In course of their analysis they have given an instance of both way steering activation using family of Doubly-Erased Werner (DEW) states\cite{new1}. Activation phenomenon considered there did not rely on testing any detection criterion in form of steering inequality. So from that perspective, the activation example\cite{new1} is comparable with that of genuine steering activation in our work. Characterization of two qubit state space based on genuine activation of steering discussed in subsec.\ref{char2} thus  encompasses a broader class of steering activation results  compared to a specific example of activation\cite{new1}. To this end one may note that for analysis made there, authors considered not only unsteearbility but also separability of the states distributed by the sources. Following that approach, incorporating entanglement content of initial unsteerable states to explore genuine activation of steering will be an interesting direction of future research. \\
\section{Data availability statement}
The manuscript has no data associated with it.
\section{Acknowledgement}
This preprint has not undergone peer review or any post-submission improvements or corrections. The Version of Record of this
article is published in The European Physical Journal D, and is available online at https://doi.org/[insert DOI]”.
\section{Appendix A}
\textit{Proof of Theorem.1}
 Both $\rho_{AB}^{'}$(Eq.(\ref{s1441})) and $\rho_{BC}^{'}$(Eq.(\ref{s1442}))
violate Eq.(\ref{mon8}). Hence, $\sum_{j=1}^3\small{\sqrt{\mathfrak{t}^2_{1jj}}},\small{\sqrt{\mathfrak{t}^2_{2jj}}}$$\leq$$1$ which imply that $|\mathfrak{t}_{kjj}|$$\leq$$1,\forall k$$=$$1,2$ and $j$$=$$1,2,3.$\\
Let $\mathcal{V}_{b_1b_2}$ denote the correlation tensor of conditional state $\rho_{AC}^{(b_1b_2)}.$
Now, two cases are considered: either one or both the parties have no non null local bloch vectors. In both the
cases, $\textmd{Tr}(\mathcal{V}_{b_1b_2}^T\mathcal{V}_{b_1b_2})$$=$$
\sum_{k=1}^3(t_{1kk}t_{2kk})^2,\,\forall b_1,b_2$$=$$0,1.$ Hence, for each of
$\mathcal{V}_{b_1b_2},$ $\sqrt{\textmd{Tr}(\mathcal{V}_{b_1b_2}^T\mathcal{V}_{b_1b_2})}$  takes the form:
\begin{eqnarray}
    \sqrt{\textmd{Tr}(\mathcal{V}_{b_1b_2}^T\mathcal{V}_{b_1b_2})}=\sqrt{\sum_{k=1}^3(t_{1kk}t_{2kk})^2}\nonumber\\
    \leq \sqrt{\sqrt{\sum_{k=1}^3}t^4_{1kk}.\sqrt{\sum_{k=1}^3}t^4_{2kk}}\nonumber\\
    \leq  \sqrt{\sqrt{\sum_{k=1}^3}t^2_{1kk}.\sqrt{\sum_{k=1}^3}t^2_{2kk}}\nonumber\\
    \leq1.
\end{eqnarray}
The second inequality holds as $|\mathfrak{t}_{kjj}|$$\leq$$1,\forall k$$=$$1,2$ and $j$$=$$1,2,3$ and the last is due to the fact that none of the initial states satisfies Eq.(\ref{mon8}).
\section{Appendix B}
\textit{Proof of Theorem.3} Here $\vec{u_1}$$=$$\vec{u_2}$$=$$\Theta.$ $\rho_{AB}$ and $\rho_{BC}$ thus have the form:
\begin{equation*}
\rho_{AB}=\small{\frac{1}{4}(\mathbb{I}_{2\times2}+\mathbb{I}_2\otimes \vec{\mathfrak{v}_1}.\vec{\sigma}+\sum_{j_1,j_2=1}^{3}w_{1j_1j_2}\sigma_{j_1}\otimes\sigma_{j_2})},
\end{equation*}
 \begin{equation*}
    \rho_{BC}=\small{\frac{1}{4}(\mathbb{I}_{2\times2}+\mathbb{I}_2\otimes \vec{\mathfrak{v}_2}.\vec{\sigma}+\sum_{j_1,j_2=1}^{3}w_{2j_1j_2}\sigma_{j_1}\otimes\sigma_{j_2})},
\end{equation*}
Let $\Lambda$(Eq.(\ref{map1})) be applied on both $\rho_{AB}$ and $\rho_{BC}$ followed by local unitary
operations(to diagonalize the correlation tensors). Let $\rho_{AB}^{(2)}$ and $\rho_{BC}^{(2)}$
denote the respective canonical forms(Eq.(\ref{map3})) of $\rho_{AB}$ and $\rho_{BC}$\cite{unsteer}:
\begin{equation}\label{map7}
\small{\rho_{AB}^{(2)}}=\small{\frac{1}{4}(\mathbb{I}_{2\times2}+\sum_{j=1}^{3}w_{1jj}^{''}\sigma_{j}\otimes\sigma_{j})},
\end{equation}
\begin{equation}\label{map8}
\small{\rho_{BC}^{(2)}}=\small{\frac{1}{4}(\mathbb{I}_{2\times2}+\sum_{j=1}^{3}w_{2jj}^{''}
\sigma_{j}\otimes\sigma_{j})},
\end{equation}
Now $\rho_{AB}^{(2)}$ and $\rho_{BC}^{(2)}$ both satisfy unsteerability criterion given by Eq.(\ref{unstr}). This in turn gives:
\begin{equation}\label{map8}
\textmd{Max}_{x_1,x_2,x_3}\sqrt{\sum_{j=1}^3}(x_jw_{kjj}^{''})^2\leq \frac{1}{2},\,\,\,\,k=1,2
\end{equation}
where $\hat{x}$$=$$(x_1,x_2,x_3)$ denotes a unit vector.
We next perform maximization over $\hat{x}$ so as to obtain a closed form of the unsteerability criterion
in terms of elements of correlation tensors of the initial states $\rho_{AB}^{(2)}$ and $\rho_{BC}^{(2)}$. \\
\textit{Maximization over unit vector $\hat{x}$:}\\
Taking $\hat{x}$$=$$(\sin(\theta)\cos(\phi),
\sin(\theta)\sin(\phi),\cos(\theta)),$ maximization problem in L.H.S. of Eq.(\ref{map8}) can be posed as:
\begin{equation}\label{map9}
\textmd{Max}_{\theta,\phi}\sqrt{A(\theta,\phi)}
\end{equation}
where,
\begin{eqnarray}\label{map10}
\small{A(\theta,\phi)}= \sin^2(\theta)(\cos^2(\phi)(w_{k11}^{''})^2&+&\nonumber\\
\sin^2(\phi)(w_{k22}^{''})^2)+\cos^2(\theta)(w_{k33}^{''})^2&&
\end{eqnarray}
Now for any $g_1,g_2$$\geq$$0,$ $\textmd{Max}_{\kappa}(g_1\cos^2(\kappa)+g_2\sin^2(\kappa))$ is $g_1$ if $g_1$$>$$g_2$ and $g_2$ when $g_2$$>$$g_1.$ This relation is used for maximizing $A(\theta,\phi).$ In order to consider all possible values of $(w_{k11}^{''})^2,$ $(w_{k22}^{''})^2$ and $(w_{k33}^{''})^2,$ we consider the following cases:\\
\textit{Case1:}$(w_{k11}^{''})^2$$>$$(w_{k22}^{''})^2$: Then $\textmd{Max}_{\phi}A(\theta,\phi)$ gives:
\begin{equation}\label{11}
    B(\theta)= \small{\sin^2(\theta)(w_{k11}^{''})^2+\cos^2(\theta)(w_{k33}^{''})^2}
\end{equation}
\textit{Subcase1:}$(w_{k11}^{''})^2$$>$$(w_{k33}^{''})^2,\textmd{i.e.,}\,(w_{k11}^{''})^2$$=$$\textmd{Max}_{j=1,2,3}
(w_{kjj}^{''})^2:$ Then $\textmd{Max}_{\theta}B(\theta)$$=$$(w_{k11}^{''})^2.$ Hence,
\begin{equation}\label{12}
\textmd{Max}_{\theta,\phi}\sqrt{A(\theta,\phi)}=|w_{k11}^{''}|.
\end{equation}
\textit{Subcase2:}$(w_{k11}^{''})^2$$<$$(w_{k33}^{''})^2,\textmd{i.e.,}\,
(w_{k22}^{''})^2$$<$$(w_{k11}^{''})^2$$<$$(w_{k33}^{''})^2:\,\textmd{Then}\,\textmd{Max}_{\theta}B(\theta)$$=$$(w_{k33}^{''})^2.$ Hence,
\begin{equation}\label{13}
\textmd{Max}_{\theta,\phi}\sqrt{A(\theta,\phi)}=|w_{k33}^{''}|.
\end{equation}

\textit{Case2:}$(w_{k11}^{''})^2$$<$$(w_{k22}^{''})^2$: Then $\textmd{Max}_{\phi}A(\theta,\phi)$ gives:
\begin{equation}\label{14}
    B(\theta)= \small{\sin^2(\theta)(w_{k22}^{''})^2+\cos^2(\theta)(w_{k33}^{''})^2}
\end{equation}
\textit{Subcase1:}$(w_{k22}^{''})^2$$>$$(w_{k33}^{''})^2,\textmd{i.e.,}\,
(w_{k22}^{''})^2$$=$$\textmd{Max}_{j=1,2,3}(w_{kjj}^{''})^2:$
Then$\quad\textmd{Max}_{\theta}B(\theta)$$=$$(w_{k22}^{''})^2.$
Hence,
\begin{equation}\label{15}
\textmd{Max}_{\theta,\phi}\sqrt{A(\theta,\phi)}=|w_{k22}^{''}|.
\end{equation}
\textit{Subcase2:}$(w_{k22}^{''})^2$$<$$(w_{k33}^{''})^2,
\textmd{i.e.,}\,(w_{k11}^{''})^2$$<$$(w_{k22}^{''})^2$$<$$(w_{k33}^{''})^2:\,\textmd{Then}\, \textmd{Max}_{\theta}B(\theta)$$=$$
(w_{k33}^{''})^2.$ Hence,
\begin{equation}\label{16}
\textmd{Max}_{\theta,\phi}\sqrt{A(\theta,\phi)}=|w_{k33}^{''}|.
\end{equation}
So, combining all cases, we get:
\begin{equation}\label{17}
\textmd{Max}_{\theta,\phi}\sqrt{A(\theta,\phi)}=\textmd{Max}_{j=1}^{3}|w_{kjj}^{''}|,\,\,k=1,2.
\end{equation}
So, the unsteerability criterion(Eq.(\ref{map8})) turns out to be:
\begin{equation}\label{18}
\textmd{Max}_{j=1,2,3}|w_{kjj}^{''}|\leq \frac{1}{2}.
\end{equation}
where $k$$=$$1,2$ correspond to states $\rho_{AB}^{(2)}$ and $\rho_{BC}^{(2)}$ respectively. So $\rho_{AB}^{(2)}$ and $\rho_{BC}^{(2)}$ and therefore $\rho_{AB}$ and $\rho_{BC}$ are unsteerable. Steerability of state remaining invariant under application of linear map(Eq.(\ref{map1})), considering the canonical forms $\rho_{AB}^{(2)}$ and $\rho_{BC}^{(2)}$ as the initial states used in the network. Depending on the output of BSM obtained by Bob(and result communicated to Alice and Charlie), the conditional states shared between Alice and Charlie are given by $\rho_{AC}^{ij},\,i,j$$=$$0,1$(see Table.\ref{table:ta15}). $\forall i,j,$ $\rho_{AC}^{ij}$ has null local blochs and diagonal correlation tensor. \\
\begin{center}
\begin{table}[htp]
\caption{ State parameters of each of the four conditional states are specified here. $\vec{X}_1,\vec{X}_2$ denote the local bloch vectors corresponding to first and second party respectively whereas $T$ denote the correlation tensor. diag($*,*,*$) stands for a diagonal matrix. Clearly each of the conditional state is in its canonical form(Eq.(\ref{map3})).}
\begin{center}
\begin{tabular}{|c|c|c|c|}
\hline
State&$\vec{X}_1$&$\vec{X}_2$&$T$\\
\hline
$\rho_{AC}^{00}$&$\Theta$&$\Theta$&diag($w_{111}^{''}w_{211}^{''},$\\
&&&$-w_{122}^{''}w_{222}^{''}$ \textbf{,} $w_{133}^{''}w_{233}^{''}$)\\
\hline
$\rho_{AC}^{01}$&$\Theta$&$\Theta$&diag($-w_{111}^{''}w_{211}^{''},$\\
&&&$w_{122}^{''}w_{222}^{''}$ \textbf{,} $w_{133}^{''}w_{233}^{''}$)\\
\hline
$\rho_{AC}^{10}$&$\Theta$&$\Theta$&diag($w_{111}^{''}w_{211}^{''},$\\
&&&$w_{122}^{''}w_{222}^{''}$ \textbf{,} $-w_{133}^{''}w_{233}^{''}$)\\
\hline
$\rho_{AC}^{11}$&$\Theta$&$\Theta$&diag($-w_{111}^{''}w_{211}^{''},$\\
&&&$-w_{122}^{''}w_{222}^{''}$\textbf{,} $-w_{133}^{''}w_{233}^{''}$)\\
\hline
\end{tabular}
\end{center}
\label{table:ta15}
\end{table}
\end{center}
Hence, for each of the conditional states, L.H.S. of Eq.(\ref{unstr}) turns out to be:
\begin{equation}\label{map19}
\textmd{Max}_{x_1,x_2,x_3}\sqrt{\sum_{j=1}^3(x_jw_{1jj}^{''}w_{2jj}^{''})^2}
\end{equation}
Following the same procedure of maximization as above, the optimal expression of the maximization problem(Eq.(\ref{map19})) is given by:
$$\textmd{Max}_{j=1,2,3}|w_{1jj}^{''}w_{2jj}^{''}|$$
Using Eq.(\ref{18}), the maximum value of Eq.(\ref{map19}) turns out to be $\frac{1}{4}.$ Each of the
four conditional states thus satisfies the unsteerability criterion(Eq.(\ref{unstr})). So if both the
initial states satisfy Eq.(\ref{unstr}) and have null local bloch vector(corresponding to first party),
then none of the conditional states generated in the network is steerable. Hence genuine activation of
steering does not occur. \\
$$\,$$
states satisfies Eq.(\ref{mon8}).
\section{Appendix C}
\textbf{\textit{Details of the numerical observation given in Sec.\ref{pur}:}} Without loss of any generality, of two initial states, let $\rho_{BC}$ has non null bloch vector corresponding to the first party, i.e.,
$\vec{u_1}$$=$$\Theta,$$\vec{u_2}$$\neq$$\Theta.$ $\rho_{BC}$ thus has the form:
\begin{equation*}
\rho_{BC}=\small{\frac{1}{4}(\mathbb{I}_{2\times2}+\vec{\mathfrak{u}_2}.\vec{\sigma}\times\mathbb{I}_{2}+\mathbb{I}_2\otimes \vec{\mathfrak{v}_2}.\vec{\sigma}+\sum_{j_1,j_2=1}^{3}w_{2j_1j_2}\sigma_{j_1}\otimes\sigma_{j_2})},
\end{equation*}
After applying $\Lambda$(Eq.(\ref{map1}))followed by local unitary operations, the canonical form $\rho_{AB}^{(2)}$ of $\rho_{AB}$ is given by Eq.(\ref{map7}) whereas that of $\rho_{BC}$ is given by:
\begin{equation}\label{map8}
\small{\rho_{BC}^{(2)}}=\small{\frac{1}{4}(\mathbb{I}_{2\times2}+\vec{\mathfrak{u}_2^{''}}.\vec{\sigma}\times\mathbb{I}_{2}
+\sum_{j=1}^{3}w_{2jj}^{''}\sigma_{j}\otimes\sigma_{j})},
\end{equation}
Now $\rho_{AB}^{(2)}$ and $\rho_{BC}^{(2)}$ both satisfy unsteerability criterion given by Eq.(\ref{unstr}). This in turn gives:
\begin{equation}\label{map28}
\textmd{Max}_{x_1,x_2,x_3}\sqrt{\sum_{j=1}^3(x_jw_{1jj}^{''})^2}\leq \frac{1}{2}
\end{equation}
and
\begin{equation}\label{map38}
\textmd{Max}_{x_1,x_2,x_3}(\vec{\mathfrak{u}_2^{''}}.\hat{x})^2+
2\sqrt{\sum_{j=1}^3(x_jw_{2jj}^{''})^2}\leq 1
\end{equation}
with $\hat{x}$$=$$(x_1,x_2,x_3)$ denoting unit vector. While the closed form of Eq.(\ref{map28}) is given by Eq.(\ref{18}) for $k$$=$$1,$ the same for Eq.(\ref{map38}) is hard to derive owing to the complicated form of the maximization problem involved in it. Now as $\rho_{BC}^{(2)}$ satisfies an unsteerability criterion(Eq.(\ref{map38})) so it is unsteerable and consequently violates Eq.(\ref{mon8}):
\begin{equation}\label{map39}
\sum_{j=1}^3(w_{2jj}^{''})^2\leq 1
\end{equation}
As discussed above, the canonical forms $\rho_{AB}^{(2)}$ and $\rho_{BC}^{(2)}$ as the initial states used in the network. Depending on Bob's output, the conditional states shared between Alice and Charlie are given by $\rho_{AC}^{ij},\,i,j$$=$$0,1$(see Table.\ref{table:ta16}).\\

\begin{center}
\begin{table}[htp]
\caption{ State parameters of each of the four conditional states are specified here. $\vec{X}_1,\vec{X}_2$ denote the local bloch vectors corresponding to first and second party respectively whereas $T$ denote the correlation tensor. diag($*,*,*$) stands for a diagonal matrix. Clearly each of the conditional state is in its canonical form(Eq.(\ref{map3})).}
\begin{center}
\begin{tabular}{|c|c|c|c|}
\hline
State&$\vec{X}_1$&$\vec{X}_2$&$T$\\
\hline
$\rho_{AC}^{00}$&($w_{111}^{''}u_{21}^{''}$\textbf{,}&$\Theta$&diag($w_{111}^{''}w_{211}^{''}$\textbf{,}\\
&$-w_{122}^{''}u_{22}^{''}$ \textbf{,} $w_{133}^{''}u_{23}^{''}$)&&$-w_{122}^{''}w_{222}^{''}$ \textbf{,} $w_{133}^{''}w_{233}^{''}$)\\
\hline
$\rho_{AC}^{01}$&($-w_{111}^{''}u_{21}^{''}$\textbf{,}&$\Theta$&diag($-w_{111}^{''}w_{211}^{''}$\textbf{,}\\
&$w_{122}^{''}u_{22}^{''}$ \textbf{,} $w_{133}^{''}u_{23}^{''}$)&&$w_{122}^{''}w_{222}^{''}$ \textbf{,} $w_{133}^{''}w_{233}^{''}$)\\
\hline
$\rho_{AC}^{10}$&($w_{111}^{''}u_{21}^{''}$\textbf{,}&$\Theta$&diag($w_{111}^{''}w_{211}^{''}$\textbf{,}\\
&$w_{122}^{''}u_{22}^{''}$ \textbf{,} $-w_{133}^{''}u_{23}^{''}$)&&$w_{122}^{''}w_{222}^{''}$ \textbf{,} $-w_{133}^{''}w_{233}^{''}$)\\
\hline
$\rho_{AC}^{11}$&($-w_{111}^{''}u_{21}^{''}$\textbf{,}&$\Theta$&diag($-w_{111}^{''}w_{211}^{''}$\textbf{,}\\
&$-w_{122}^{''}u_{22}^{''}$ \textbf{,} $-w_{133}^{''}u_{23}^{''}$)&&$-w_{122}^{''}w_{222}^{''}$\textbf{,} $-w_{133}^{''}w_{233}^{''}$)\\
\hline
\end{tabular}
\end{center}
\label{table:ta16}
\end{table}
\end{center}

Let us consider $\rho_{AC}^{00}.$ Using state parameters(Table.\ref{table:ta16}) of $\rho_{AC}^{00},$ L.H.S. of Eq.(\ref{unstr}) becomes:
\begin{eqnarray}\label{map40}
\textmd{Max}_{x_1,x_2,x_3}((x_1\mathfrak{u}_{21}^{''}w_{111}^{''}-x_2\mathfrak{u}_{22}^{''}w_{122}^{''}
+x_3\mathfrak{u}_{23}^{''}w_{133}^{''})^2&+&\nonumber\\
\sqrt{\sum_{j=1}^3(x_jw_{1jj}^{''}w_{2jj}^{''})^2})&&,
\end{eqnarray}
where $\mathfrak{u}_{21}^{''},\mathfrak{u}_{22}^{''},\mathfrak{u}_{23}^{''}$ are the components of real valued vector bloch vector $\vec{\mathfrak{u}_{2}^{''}}.$ In Eq.(\ref{map40}), maximization is to be performed over $x_1,x_2,x_3$ whereas the state parameters are arbitrary. Now the expression in Eq.(\ref{map40}) is numerically maximized over all the state parameters involved and also $x_1,x_2,x_3$ under the following restrictions:\\
\begin{itemize}
  \item $w_{111}^{''}$$\leq$$ \frac{1}{2}$
  \item $w_{122}^{''}$$\leq$$ \frac{1}{2}$
  \item $w_{133}^{''}$$\leq$$ \frac{1}{2}$
  \item $\sum_{j=1}^3(w_{2jj}^{''})^2$$\leq$$ 1.$
\end{itemize}
While the first three restrictions are due to the unsteerability of $\rho_{AB}^{(2)},$ i.e., given by  Eq.(\ref{18}) for $k$$=$$1,$ the last restriction is provided by Eq.(\ref{map39})(a consequence of unsteerability of $\rho_{BC}^{(2)}$). Maximum value of the above maximization problem(Eq.(\ref{map40})) turns out to be $0.75$, corresponding maxima(alternate maxima exists) given by $w_{111}^{''}$$=$$0.5,$ $w_{122}^{''}$$=$$0.454199,$ $w_{133}^{''}$$=$$0.46353,$  $w_{211}^{''}$$=$$-1,$ $w_{222}^{''}$$=$$0,$ $w_{233}^{''}$$=$$0,$ $\mathfrak{u}_{21}^{''}$$=$$1,$ $\mathfrak{u}_{22}^{''}$$=$$0,$ $\mathfrak{u}_{23}^{''}$$=$$0,$ $x_1$$=$$1,$ $x_2$$=$$0$ and $x_3$$=$$0.$ Maximum value less than $1$ implies that the original maximization problem(Eq.(\ref{map40})), where maximization is to be performed only over $x_1,x_2,x_3$(for arbitrary state parameters) under the above restrictions(resulting from unsteerability of $\rho_{AB}^{(2)},\rho_{BC}^{(2)}$), cannot render optimal value  greater than $1.$ Consequently conditional state $\rho_{AC}^{00}$ satisfies the unsteerability criterion(Eq.(\ref{unstr})) and is therefore unsteerable. So, in case Bob's particles get projected along $|\phi^{+}\rangle,$ genuine activation of steering does not occur in the linear network. In similar way, considering, other three conditional states, it is checked that the unsteerability criterion(Eq.(\ref{unstr})) is satisfied in each case. Genuine activation of steering is thus impossible for all possible outputs of Bob. Hence when one of the initial states has null local bloch vector corresponding to first party, genuine activation of steering does not occur.
\section{Appendix D}

\begin{center}
\begin{table}[htp]
\caption{Bloch matrix parameters of each of the four conditional states generated in the linear swapping
network using $\Omega_3,\Omega_4$(Eq,(\ref{unstr1}) as initial states are specified here. $\vec{U},\vec{V}$ denote the
local bloch vectors corresponding to first and second party respectively whereas $\textbf{T}$ denote
the correlation tensor. diag($*,*,*$) stands for a diagonal matrix. Now, for each conditional state,
$\textbf{T}$ being in diagonal form and $\vec{V}$$=$$\Theta,$ for all $i,j,$ $\rho_{AC}^{ij}$ is in
its canonical form(Eq.(\ref{map3})).}
\begin{center}
\begin{tabular}{|c|c|c|c|}
\hline
State&$\vec{U}$&$\vec{V}$&$\textbf{T}$\\
\hline
$\rho_{AC}^{00}$&$(0,0,0.98107)$&$\Theta$&diag($0.0729052,-0.0729052,$\\
&&&$0.0128697$)\\
\hline
$\rho_{AC}^{01}$&$(0,0,0.98107)$&$\Theta$&diag($-0.0729052,0.0729052,$\\
&&&$0.0128697$)\\
\hline
$\rho_{AC}^{10}$&$(0,0,0.907448)$&$\Theta$&diag($0.0729052,0.0729052,$\\
&&&$-0.0128697$)\\
\hline
$\rho_{AC}^{11}$&$(0,0,0.907448)$&$\Theta$&diag($-0.0729052,-0.0729052,$\\
&&&$-0.0128697$)\\
\hline
\end{tabular}
\end{center}
\label{table:ta17}
\end{table}
\end{center}

\end{document}